\documentclass[12pt,a4paper]{article} 	
\pdfoutput=1
\usepackage{jheppub}
\usepackage{graphicx}
\usepackage{amssymb}
\usepackage{amsmath}
\usepackage{bbold}
\usepackage{tikz}
\usepackage{array}

\allowdisplaybreaks
\newcommand{\be}{\begin{equation}}
\newcommand{\ee}{\end{equation}}
\newcommand{\bea}{\begin{eqnarray}}
\newcommand{\eea}{\end{eqnarray}}
\def\bse{\begin{subequations}}
\def\ese{\end{subequations}}
\newcommand{\IR}{\mathbb{R}} 
 
\def\IZ{\relax\ifmmode\hbox{Z\kern-.4em Z}\else{Z\kern-.4em Z}\fi}

\newcommand{\non}{\nonumber \\}

\newcommand{\Li}{\mathrm{Li}}
\newcommand{\Tad}{\mathrm{Tad}}
\newcommand{\Bub}{\mathrm{Bub}}
\newcommand{\Sun}{\mathrm{Sun}}

\def\half{\frac{1}{2}} 
\def\del{{\partial}} 

\def\presub{\vspace{.5cm} \noindent}

\def\bi{\begin{itemize}} \def\ei{\end{itemize}}

\def\({\left(} \def\){\right)}
\def\[{\left[} \def\]{\right]}

\def\lam{\lambda} \def\eps{\epsilon}
\def\al{\alpha}

\title{The propagator seagull: general evaluation of a two loop diagram}
\author{Barak Kol and Ruth Shir \\
{\it The Racah Institute of Physics, The Hebrew University of Jerusalem,}\\ \it{Jerusalem 91904, Israel}\\
{\tt  barak.kol, ruth.shir@mail.huji.ac.il}
}

\abstract{We study a two loop diagram of propagator type with general parameters through the Symmetries of Feynman Integrals  (SFI) method. We present the SFI group and equation system, the group invariant in parameter space and a general representation as a line integral over simpler diagrams. We present close form expressions for three sectors, each with three or four energy scales, for any spacetime dimension $d$ as well as the $\epsilon$ expansion. We determine the singular locus and the diagram's value on it.}

\begin{document}
\maketitle

\section{Introduction}

Analytic calculation of Feynman diagrams with several mass scales is a challenge, but it is important for higher loop corrections to Standard Model/Core Theory\footnote{
The term Core Theory was advocated in \cite{CoreTheory} to supersede the term Standard Model.} 
observables involving several different particles, and it is of intrinsic interest to Quantum Field Theory. 

The Symmetries of Feynman Integrals method (SFI) \cite{SFI}, see also developments in \cite{locus,bubble,VacuumSeagull,minors,diameter,kite}, reduces the diagram to its value at some allowed and more convenient base point in parameter space, namely the space $X$ of masses and kinematical invariants, plus a line integral in $X$ over simpler diagrams (with one edge contracted).  

This method is close in spirit to the Integration By Parts (IBP) method \cite{Chetyrkin:1981qh} and to the Differential Equations (DE) method \cite{Kotikov:1990kg, Remiddi:1997ny,GehrmannRemiddi1999,Caffo:1998yd}, see also the textbooks \cite{SmirnovBooks}. The main new feature of SFI is recognizing a continuous group $G$ which is associated with the Feynman Integral and identifying its action in parameter space, thereby leading to the above-mentioned reduction. See appendix \ref{sec:relation} for an extended discussion of the relation with IBP and DE. 

\begin{figure}
\centering \noindent
\includegraphics[width=10cm]{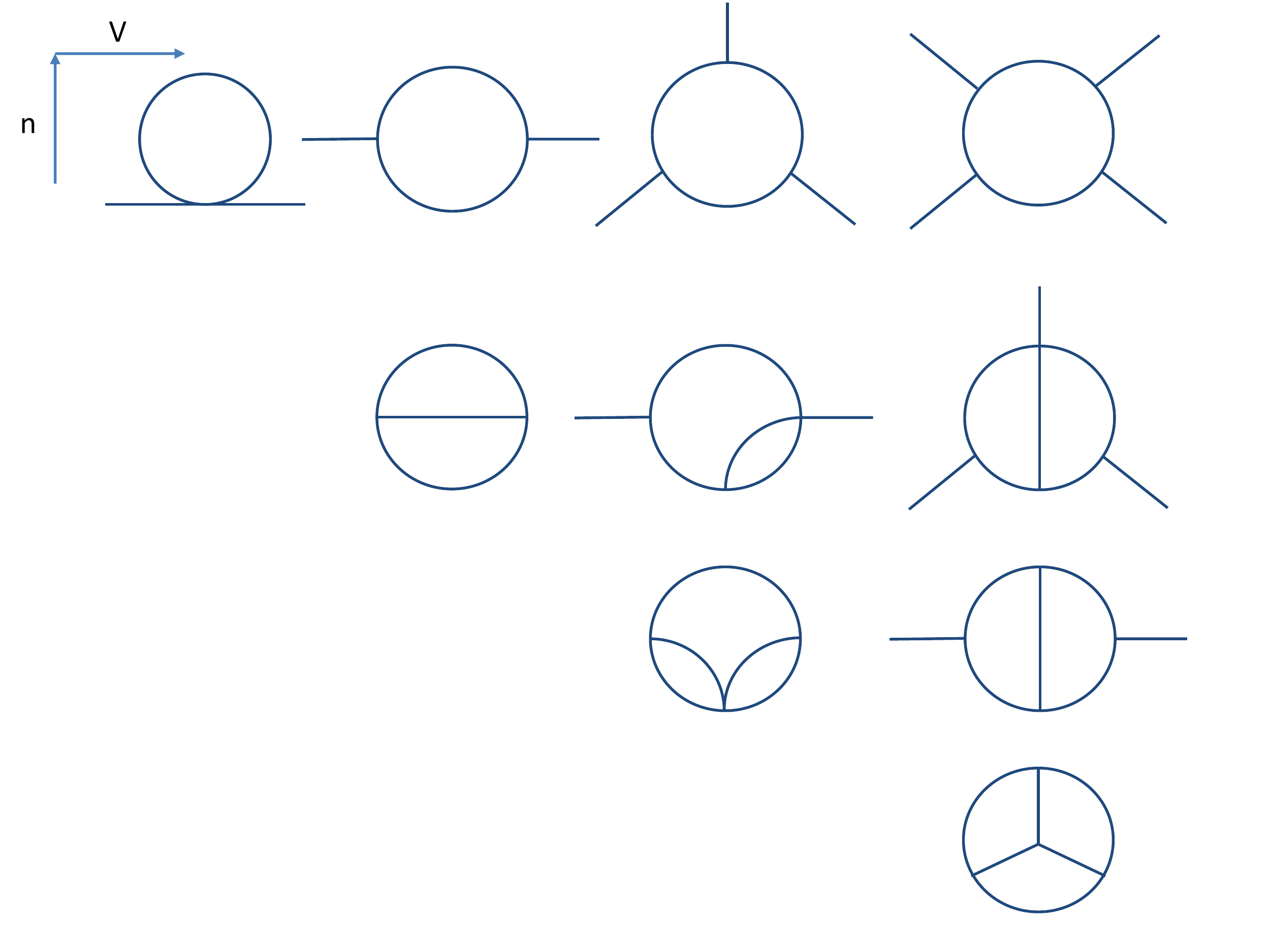} 
\caption[]{Hierarchy of diagrams according to edge contraction.  Each column has diagrams of fixed number of vertices so that necessary sources for each diagram are always on its left.  The propagator seagull is on the third column from the left, second from bottom.}
\label{fig:DiagHierarchy} \end{figure}

SFI offers a new approach to the evaluation of Feynman diagrams and it is important to demonstrate it and further develop it through application to specific diagrams. The method suggests to partially order all diagrams according to edge contraction as shown in Fig.~\ref{fig:DiagHierarchy}. The leftmost column consists of the tadpole which is immediate to evaluate through integration of Schwinger parameters. The next column includes the diameter and bubble diagrams which were treated through SFI in \cite{diameter,bubble} where novel derivations for their known values were described. The vacuum seagull is in the bottom of the third column, and its analysis through SFI enabled a a novel evaluation of a sector with three mass scales \cite{VacuumSeagull}. The kite diagram, which is on the fourth row from the left, second from bottom, was analyzed through SFI in \cite{SFI} and it was possible to identify a locus in parameter space where it reduces to a linear combination of simpler diagrams, thereby maximally generalizing the massless case, studied in \cite{Chetyrkin:1981qh}, and the application of the diamond rule of \cite{DiamondRule}.

In this paper we analyze a diagram which we call the propagator seagull. It is second from bottom on the third column from the left in Fig.~\ref{fig:DiagHierarchy}. It is a two-loop diagram of a propagator type which can be gotten from the vacuum seagull by cutting open one of its two parallel loops, thereby explaining its name. Being on the third column, the propagator seagull appears as a source for more complicated diagrams, such as the tetrahedron and the kite (fourth column, bottom two). In addition we shall see it is special in providing a first example where the group orbits are not open in parameter space, but rather are of co-dimension 1. 

\cite{Tarasov1997} studied all two-loops diagrams of propagator type and reduced the case of arbitrary indices (powers of propagators) to the kite, the propagator seagull and the sunrise (the master integrals). \cite{Bauberger1994}, section 4.3, expressed the general case in the small $p^2$ limit in terms of a straightforward quadruple sum which is a generalized  hypergeometric series. In \cite{Davydychev:2000na} the propagator seagull was considered with one mass scale but with arbitrary indices using the Mellin-Barnes method \cite{Boos:1990rg}. The propagator seagull with two mass scales was calculated using dispersion relations in \cite{Scharf:1993ds}, using dispersion relations with Mellin-Barnes method in \cite{Bauberger:1994by}, using an $\epsilon$-expansion for the differential equation method in \cite{Jegerlehner:2003sp} and using Mellin-Barnes in \cite{Bytev:2011ks}. Other two mass scale sectors were calculated in \cite{Gray:1990yh} via Integration By Parts reduction and recalculated in \cite{Martin:2005qm}. For three mass scales an analytic expression for any dimension involving Appell hypergeometric functions was given in \cite{Bauberger:1994by} and in \cite{Scharf:1993ds} and \cite{Martin:2003it} in an $\epsilon$ expansion. With four mass scales the $\epsilon^0$ term was calculated through differential equations in \cite{Martin:2003qz}. 

The paper is organized as follows. We start by deriving the SFI equation system and the associated group in section \ref{sec:eq_system}. Next, in section \ref{g_orbits} we obtain the group orbits in parameter space and the associated group invariant. Section \ref{sec:gen_soln} presents a general formula for a reduction to a base point (where two masses are put to zero) plus a line integral over simpler diagrams. Section \ref{sec:sectors} contains the evaluation of the line integral for several sectors with 3 and 4 energy scales as well as the $\epsilon$ expansion. Section \ref{al} determines the singular locus where the diagram simplifies to a linear combination of simpler diagrams (rather than a line integral thereof) and presents the corresponding evaluations. We summarize our results and disucss them in section \ref{sec:summary}. Finally an appendix  provides  a useful collection of diagram evaluations and definitions of special functions.

\section{SFI group and equation system}
\label{sec:eq_system}

\begin{figure}[t]
\begin{center}
\includegraphics[scale=0.35]{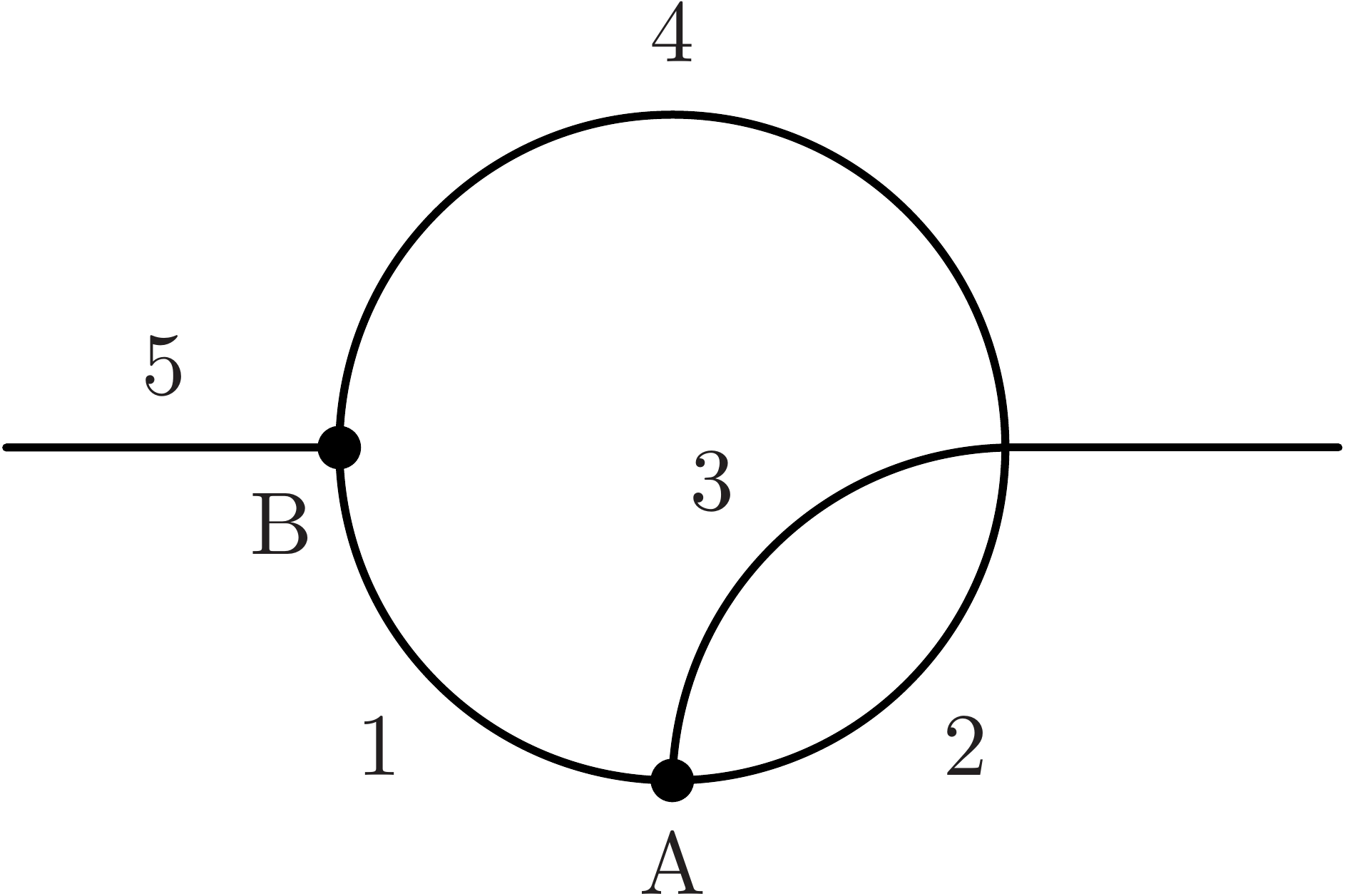}
\caption{The propagator seagull}
\label{propagator_seagull}
\end{center}
\end{figure}

In this paper we will consider the two-loop propagator type diagram shown in Figure~\ref{propagator_seagull} and given by the integral
\begin{equation}\label{loop_int}
I(x_1,x_2,x_3,x_4,x_5)=\int \frac{d^dl_1\, d^dl_2}{(l_1^2-x_1)(l_2^2-x_2)((l_1+l_2)^2-x_3)((l_1+p)^2-x_4)}
\end{equation}
where the parameter space is \be 
X=\{(x_1,\, x_2,\, x_3,\, x_4, x_5 ) = \( m_1^2,\, m_2^2,\, m_3^2,\, m_4^2 ,\, p^2\) \}. 
\ee 
The discrete symmetry group is a reflection exchanging propagators 2 and 3, namely \be
\Gamma= \IZ_2 ~.
\ee 

\presub {\bf SFI method}. Before we proceed to obtain the SFI equation system let us give a short review of the SFI method. The integral (\ref{loop_int}) is invariant under transformations of the integration variables $l_A$, and in particular under infinitesimal linear variations of the form
\begin{eqnarray}
l_A &\to& l_A +\epsilon_{AB} l_B +\epsilon_{A}p \label{trans1} 
\end{eqnarray}
with $A,B=1,2$ and $\epsilon_{AB},\epsilon_A$ are small.\footnote{The notation $\epsilon_{AB}$ should not be confused with the 2d  volume tensor.} When applied to the integral (\ref{loop_int}) such variations induce the following equations
\begin{eqnarray*}
0&=&\int d^dl_1\, d^dl_2\, \left(\frac{\partial}{\partial l_A} l_B\, \tilde{I}\,\right) \\
0&=&\int d^dl_1\, d^dl_2\, \left( \frac{\partial}{\partial l_A} p\, \tilde{I}\,\right)
\end{eqnarray*}
with $\tilde{I}$ the integrand. 

Another useful equation is obtained by the variation 
\be
p \to p+\epsilon_p p \label{trans3}
\ee
which does not leave the integral invariant but nevertheless gives the following useful identity\footnote{Equivalently $2 p^2 \del_{p^2} =p^\mu \del_{p^\mu}$ in the presence of a single independent external momentum.}
\begin{equation}
2 p^2 \frac{\partial}{\partial p^2}I=\int d^dl_1\, d^dl_2\,p \frac{\partial}{\partial p} \, \tilde{I}\,~.
\end{equation}
One can consider in this way variations of the form (\ref{trans1}, \ref{trans3}) with 7 $\eps$ parameters:
\begin{equation}
\delta \begin{pmatrix} l_1 \\l_2\\p \end{pmatrix}=\begin{pmatrix} \epsilon_{11} & \epsilon_{12} & \epsilon_{1} \\ \epsilon_{21} & \epsilon_{22} & \epsilon_{2} \\ 0 & 0 & \epsilon_p \end{pmatrix}\begin{pmatrix} l_1 \\l_2\\p \end{pmatrix}~. \label{trans_pre}
\end{equation}
We define the space of quadratics (or current scalar products)
\be
Q:={\rm Sp} \{ l_1^2,\, l_1 \cdot l_2,\, l_2^2, \,  l_1\cdot p,\, l_2 \cdot p,\, p^2 \}
\ee
and the variation (\ref{trans_pre}) induces a variation on $Q$. We record the following variations which will be needed shortly, keeping terms up to first order in $\epsilon$:
\bea
\delta(l_1^2)&=&2\epsilon_{11} l_1^2+2\epsilon_{12}l_1\cdot l_2+2\epsilon_1 l_1\cdot p \non 
\delta(l_1\cdot l_2)&=&\epsilon_{21}l_1^2+\epsilon_{12}l_2^2+(\epsilon_{11}+\epsilon_{22})l_1\cdot l_2+\epsilon_2 l_1\cdot p+\epsilon_1 l_2\cdot p \non
\delta(l_2^2)&=&2\epsilon_{22}l_2^2+2\epsilon_{21}l_1\cdot l_2+2\epsilon_{2}l_2\cdot p \non 
\delta(l_1\cdot p)&=& \epsilon_{11}l_1\cdot p+\epsilon_{12}l_2\cdot p+\epsilon_{1}p^2 . \label{deltaQ}
\eea

In order to vary the integral we write it as  \be
I(x)=\int d^dl_1\,d^dl_2 \tilde{I}(l_1,l_2,p,x)
\ee
with the integrand $\tilde{I}(l_1,l_2,p,x)$ given by
\be
\tilde{I}(l_1,l_2,p,x)=\prod_{i=1}^4 \frac{1}{P_i-x_i}, \quad P_i := \{l_1^2,\, l_2^2,\, (l_1+l_2)^2,\, (l_1+p)^2\}.
\ee
 $P_i$ is the list of squared internal currents (propagators without the masses). 

The variation (\ref{trans_pre}) induces the following transformation on $\tilde{I}$:
\be
\delta \tilde{I}=-\prod_{i=1}^4 \frac{\delta P_i}{P_i-x_i}\tilde{I}.
\ee
 $\delta P_i$  can be gotten from the induced transformations $\delta Q$ (\ref{deltaQ}). For example
\be
\delta (l_1+l_2)^2=\delta(l_1^2)+2\delta (l_1\cdot l_2)+\delta (l_2^2).
\ee

A variation generates a differential equation for $I$ exactly when $\delta P_i$ belongs to the space of squares \be
S:={\rm Sp} \{l_1^2,\, l_2^2,\, (l_1+l_2)^2,\, (l_1+p)^2,\, p^2 \}
\ee
namely, $S$ is spanned by both squares of internal currents (propagators) and scalar products of external ones (kinematical invariants). 

Since the space of quadratics $Q$ is 6d and the space of squares $S$ is 5d there is a single irreducible scalar product (ISP), or irreducible numerator, which we choose to be $l_2 \cdot p$. Variations which avoid the generation of the irreducible numerator, or equivalently, those which preserve $S$, generate the differential equations of interest. By looking at (\ref{deltaQ}) we see that the variations $\epsilon_1,\epsilon_2$ and $\epsilon_{12}$ will generate a numerator $l_2\cdot p$ and are therefore set to zero.


In this way we find that the group $G$ of linear transformations on the space of loop and external momenta can be represented by the following variations 
\begin{equation}
\delta \begin{pmatrix} l_1 \\l_2\\p \end{pmatrix}=\begin{pmatrix} \epsilon_{11} & 0 & 0 \\ \epsilon_{21} & \epsilon_{22} & 0 \\ 0 & 0 & \epsilon_p \end{pmatrix}\begin{pmatrix} l_1 \\l_2\\p \end{pmatrix}~. \label{trans}
\end{equation}

\presub {\bf SFI equation system}. The variation (\ref{trans}) translates into operating on the integrand by either one of the following variations $\frac{\partial}{\partial l_1}l_1$, $\frac{\partial}{\partial l_2}l_2$, $\frac{\partial}{\partial l_2}l_1$, or $\frac{\partial}{\partial p}p$, which in turn give a system of 4 partial differential equations in parameter space of the form:
\begin{equation}
c^a\,I(x)-(Tx)^a_i\frac{\partial}{\partial x_i}I(x)=J^a(x) \label{PDEs}
\end{equation}
where $a=1,\dots,4$ enumerates the equations and $i=1,\dots ,5$ enumerates the parameters. $c^a$ are constants which could depend on the number of spacetime dimensions $d$; $Tx$ is a $4\times 5$ matrix with entries linear in the $x$'s and $J^a(x)$ are linear combinations of Feynman diagrams which arise from $I(x)$ by contracting one propagator. 

We choose to present the equation system in the basis defined by the following variations \be
\begin{pmatrix}
 E^1 \\
 E^2 \\
 E^3 \\
 E^4
\end{pmatrix}
:=
\begin{pmatrix}
\frac{\partial}{\partial l_1}l_1+\frac{\partial}{\partial l_2}l_2+\frac{\partial}{\partial p}p \\
\frac{\partial}{\partial l_2} l_2 \\
\frac{\partial}{\partial l_2}(l_1+l_2) \\
\frac{\partial}{\partial p} p 
\end{pmatrix}. \label{def:E} \ee
In it (\ref{PDEs}) reads
\begin{equation} \label{c}
c = \begin{pmatrix}
2d-8 \\ d-3 \\ d-3 \\ -1 
\end{pmatrix},
\quad
J(x) = \begin{pmatrix}
0 \\
\partial^3 (O_2 - O_1)I \\
\partial^2 (O_3 - O_1)I\\
-\partial^4 O_1I
\end{pmatrix}
\end{equation}
\begin{equation} \label{Tx}
Tx= 2 \begin{pmatrix}
 x_1 & x_2 & x_3 & x_4 &  x_5 \\ 
0 & x_2 & s^1_A  & 0 & 0 \\
0  & s^1_A & x_3 & 0 & 0 \\
0 & 0 & 0 & s^1_B & x_5 
\end{pmatrix}~.
\end{equation}
This basis is adapted to the reflection symmetry of the diagram in the sense that equations 1 and 4 are invariant under reflection, while 2 and 3 are exchanged.

The notation above is defined as follows. The $s$ variables are defined for any trivalent vertex $v$ by \be
 s^i_v := (x_j + x_k - x_i)/2
\label{def:si}
\ee
where $i,j,k$ are the three edges which meet at $v$ (see also \cite{diameter,kite}). The propagator seagull has two trivalent vertices: $A := (123),\, B:=(145)$, as shown in the fig. \ref{propagator_seagull} and hence   \bea
s^1_A &:=& (x_2 +x_3 -x_1)/2 \non
s^1_B &:=& (x_4 +x_5 -x_1)/2 ~. 
\eea
For $i=2,\dots,5$ propagator $i$ belongs exactly to a single trivalent vertex (either $A$ or $B$) and hence the subscript denoting the vertex may be omitted. For example \be
s^5 \equiv s^5_B := (x_1 +x_4 -x_5)/2 ~.
\ee
The $O_i$ operators $i=1,\dots,5$ which appear in $J^a$ denote the diagram gotten by omitting, or contracting, the $i$'th propagator. Note however, that $O_4$ does not appear in the SFI equations. Accordingly, the possible topologies for the degenerations are 
\begin{equation}\label{sources}
\mathrm{Degen}\left[ \raisebox{-12pt}{\includegraphics[scale=0.3]{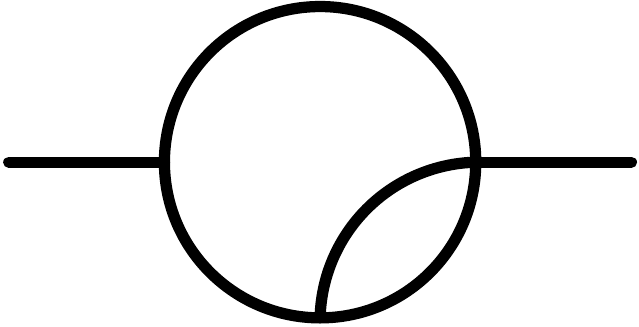}}\right]=\left\{\raisebox{-12pt}{\includegraphics[scale=0.3]{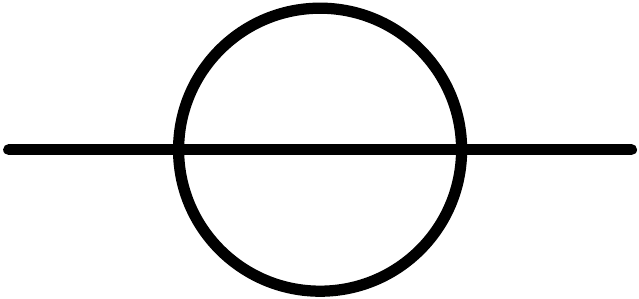}}\, , \; \raisebox{-7pt}{\includegraphics[scale=0.2]{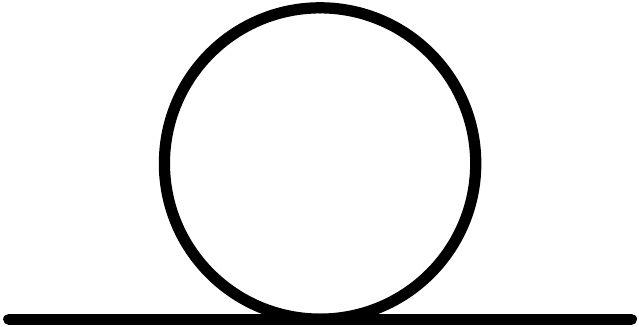}} \times \raisebox{-12pt}{\includegraphics[scale=0.3]{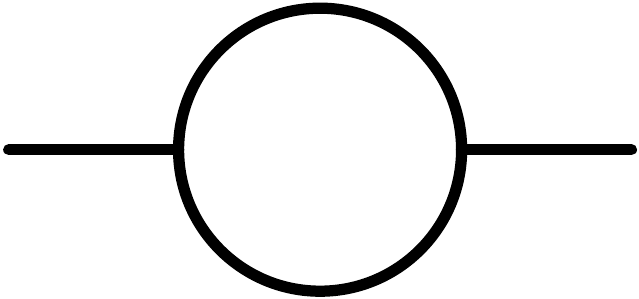}} \right\},
\end{equation}
namely, the sunrise and the tadpole times the bubble.
Finally, the derivatives are defined as $\del^i:=\del/\del x_i$.

For later use we define the Heron / K\"all\'en invariant \be
 \lam := x^2 + y^2 + z^2 - 2\, x\, y- 2\, x\, z - 2\, y\, z
\ee
 and for each trivalent vertex $v$ \be
  \lam_v := \lam(x_i, x_j,  x_k)
\ee
so that in particular
\bea 
 \lam_A &:=& \lam (x_1, x_2, x_3) \non
 \lam_B  &:=&  \lam (x_1, x_4, x_5) ~.\label{def:lam}
\eea 
Motivation and geometric interpretation for the definitions of $\lam$ and $s$ can be found in \cite{diameter}.

\newpage
\presub {\bf Comments}. 

{\bf Relation with vacuum seagull.} The group $G$ given by (\ref{trans}) is of the form \be
G=G_{vac} \cap T_{2,1} = 
\begin{pmatrix} * & 0 & 0 \\ * & * & 0 \\ * & 0 & * \end{pmatrix}  
\cap
\begin{pmatrix} * & * & * \\ * & * & * \\ 0 & 0 & * \end{pmatrix} = 
\begin{pmatrix} * & 0 & 0 \\ * & * & 0 \\ 0 & 0 & * \end{pmatrix}
\ee
where $G_{vac}$ is the group for the vacuum closure of the diagram, namely the vacuum seagull \cite{VacuumSeagull}, $T_{2,1}$ is a triangular block matrix describing the variations (\ref{trans1},\ref{trans3}) and a $*$ denotes an arbitrary entry. This is in agreement with the general result for a diagram with 2 or 3 external legs, see \cite{bubble} eq. (2.16) and below.

{\bf Degeneracy of obstructions.} By obstructions we mean terms which may appear in a loop current variation of the form (\ref{trans1}) and prevent it (or obstruct it) from producing a differential equation for $I$ defined in (\ref{loop_int}). Such terms are of the form $N_i \del_{x_j}$, where $N_i$ is an ISP, or irreducible numerator. 

$G$ has the property that its dimension is larger by 1 from the minimum dimension gotten by considering possible obstructions, namely  \be
{\rm dim} (Obst) = {\rm dim}(Num) \cdot {\rm dim}(Prop) = 1 \cdot 4 = 4 
\ee
while \bea
 {\rm dim}(G) &=& 4 \non
 			&\ge& {\rm dim}(T_{2,1}) - {\rm dim}(Obst) = 7- 4 =3 .
\eea

This means that not all of the possible obstructions are actually generated by the variations. We find that the actual obstructions can be characterized as those which do not include $\del/\del x_1$, or equivalently, those which annihilate $x_1$. 
In other words none of the 7 variations (\ref{trans1},\ref{trans3}) produces the term $N\, \del_{x_1} $ where $N=l_2 \cdot p$ is the irreducible scalar product. 
Precisely the same comment applies to the vacuum seagull.

{\bf  Forbidden terms.} In the presence of external legs we consider all variations except for those of the form $\del_p\, l$. Hence, terms of the form \be
x\,  \frac{\del}{\del (p_r \cdot p_s)}
\ee
cannot appear in the equations, where $x$ is any mass-squared. We note that by taking the equation system for the vacuum seagull and keeping the maximal subspace which does not include such forbidden terms one reproduces $G$.


\section{Group orbits}
\label{g_orbits}
The matrix $Tx$ in the SFI equation system (\ref{PDEs}-\ref{Tx}) defines an action of the SFI group $G$ on parameter space $X$. In this section we study the geometry of the $G$ orbits, which are the characteristic surfaces for the SFI equation set. 

We have 4 equations in a 5d parameter space, hence the group orbits are at most 4d. According to the method of maximal minors \cite{minors}, to find the group orbits we should compute the 4-minors $M^i$ gotten from $Tx$ (\ref{Tx}) by omitting column $i$, taking the determinant and multiplying by an alternating sign (\cite{minors} provides a full introduction and the motivation). Equivalently, it is calculated through the 5-dimensional completely anti-symmetric tensor
\be
 M^i =\epsilon^{i\,i_1\,i_2\,i_3\,i_4}(Tx)^1_{i_1}(Tx)^2_{i_2}(Tx)^3_{i_3}(Tx)^4_{i_4}.
\ee
One finds
\be
M^i\,dx_i=4\lambda_A\,Inv  \label{def:4minor}
\ee
with $\lambda_A$ defined in (\ref{def:lam}) and where the 1-form $Inv$ related to invariants of the SFI group is given by
\be  \label{def:inv}
Inv := Inv^i\, dx_i = x_5\, s^5\, dx_1-  x_1\, x_5\, dx_4 +  x_1\, s^1_B\, dx_5 ~.
\ee

Since $M^i$ are not identically zero we conclude that generically a $G$-orbit is indeed 4d, or equivalently,
\be
{\rm codim}(\text{ $G$ - orbits}) = 1.
\label{co-dim_g_orbit}
\ee

The propagator seagull is our first example for a diagram with non-zero co-dimension (previous examples were codimension 0 and include the diameter \cite{diameter}, the bubble \cite{bubble}, the vacuum seagull \cite{VacuumSeagull} and the kite \cite{kite}). 

We identify the common factor in (\ref{def:4minor}) to be  \be
 S =4\, \lam_A
\label{def:S}
\ee
where $S$ stands for ``singular locus'' since when $S=0$ the maximal minors all vanish and hence it defines a singular group orbit (of lower dimension). This orbit will be studied in section \ref{al}.


Note that $Inv$ involves only the parameters $x_1, x_4, x_5$ (associated with vertex $B$) and hence it can be considered as a 1-form in 3d space. This means that $m_2, m_3$ can be varied without leaving the $G$-orbit, and in particular \be
 \text{ every } G\text{-orbit includes a point where } m_2=m_3=0 ~.
\label{base}
\ee

The group orbits are co-dimension 1 in the parameter space $X$ (\ref{co-dim_g_orbit}) and hence the group possesses a single invariant, which we denote by $\phi$. Points in parameter space with different values of $\phi$ cannot be related to each other through the SFI equations. We set to determine $\phi$ with the help of the invariant 1-form $Inv$ which annihilates the tangent bundle to $G$-orbits, namely $(Tx)^a_i \, Inv^i=0$ for $a=1,\dots,4$. Hence $Inv$ is dual (``perpendicular'') to surfaces of constant $\phi$, and therefore must be proportional to the differential of $\phi$, namely \be
Inv = f\, d\phi
\label{Inv_phi} 
\ee
for some function $f$. 

In order to determine $\phi$ we may first solve for $f$ through \be
dInv=d \log f \wedge Inv ~. 
\label{dInv}
\ee
We find $f=(x_1 x_5)^{3/2}$, and substituting back into (\ref{Inv_phi}) and integrating we obtain the $G$-invariant \be
\phi=\frac{s^4}{\sqrt{x_1x_5}} ~.
\label{phi_value}
\ee

\presub {\bf Comments}.

{\bf Alternative derivation.} (\ref{Inv_phi}) implies not only (\ref{dInv}) but also \be
Inv \wedge d\phi =0
\ee
which can be used as an equation for $\phi$, in fact a system of linear first order partial differential equations, to be solved by the method of characteristics, thereby providing an alternative derivation.

{\bf Freedom of definition of invariant.} Naturally any function of $\phi$ could serve equally well as the $G$ invariant. For instance, given that $\lam_B=(s^4)^2-x_1 x_5$ we could choose any of the following alternative invariants \be
 \phi^2 \equiv \frac{(s^4)^2}{x_1x_5}, ~\frac{\lam_B}{x_1 x_5}, ~ \frac{(s^4)^2}{\lam_B} ~.
 \ee

\begin{figure}[t]
\begin{center}
\includegraphics[scale=0.3]{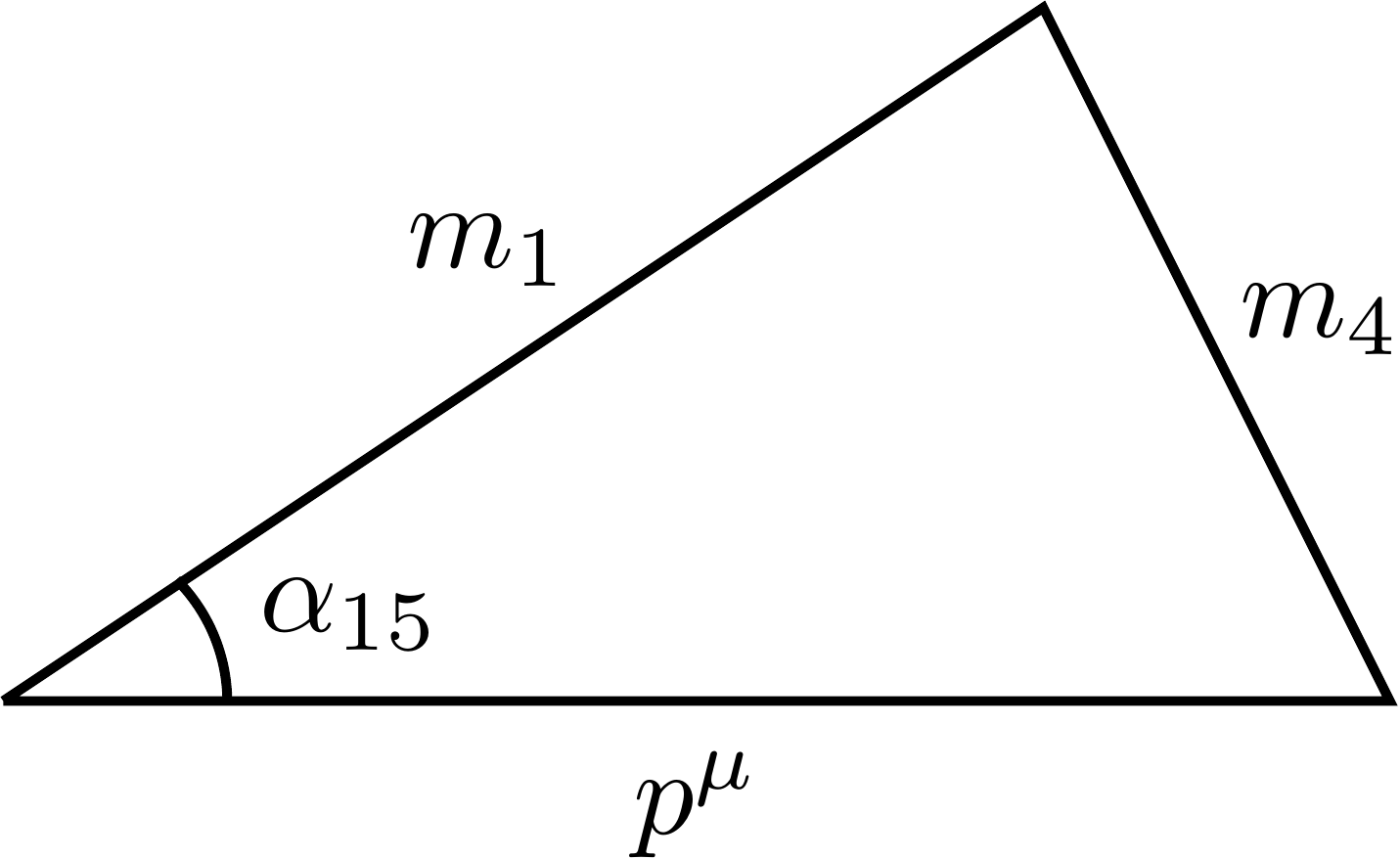}
\caption{The invariant $\phi$ can be interpreted geometrically as a function of the angle $\al_{15}$ in the shown triangle.}
\label{Inv_geom}
\end{center}
\end{figure}

{\bf Geometric interpretation of $\phi$.} In terms of the dual on-shell diagram (see \cite{bubble,kite}) the invariant is related to a triangle where $p^{\mu}$ is one of the edges and the other edges are of length $m_1$ and $m_4$. The signature of the triangle plane is determined by the data, and the triangle always exists (when the data violates the triangle inequality the triangle is defined in a Lorentzian signature plane). More precisely, $\phi$ is a function of $\al_{15}$, the angle between $m_1$ and $p$, see Fig.~\ref{Inv_geom}. For instance, if the triangle  is Euclidean then $\phi=\cos \al_{15}$. This means that $\al_{15}$ is $G$-invariant.

\section{General solution}
\label{sec:gen_soln} 

In this section we reduce the integral at a general value of the parameters to a line integral over simpler diagrams. Towards this goal we perform two intermediate steps, determining the constant-free invariants and the homogeneous solution. (Note that the maximal cut is known to be a homogeneous solution of the differential equations \cite{{MaximalCut},{MaximalCut1},{MaximalCut2}}).

\presub {\bf Constant free equations}. The constant-free equation system is defined as a linear subspace of the original equation system (with $d$ independent coefficients) such that the vector of constants vanishes, $c^a_{cf}=0$. The constant-free matrix of group generators associated with (\ref{PDEs}) is given by \be
\begin{pmatrix}
E^2 +  E^3 + 2 E^4 - E^1 \\
E^2-E^3
\end{pmatrix}  \to
\(Tx\)_{cf} = 2
\begin{pmatrix}
-x_1 	& s^1_A	& s^1_A	 & x_5-x_1 & ~ x_5 \\
 0 	& s^3 	& -s^2	 & 0 & 0
\end{pmatrix}. \ee

This defines a two dimensional group $G_{cf}$ within $G$. It must have $5-2=3$ invariants. By substituting ansatze which are either linear or quadratic in the $x$'s we find  the following constant-free invariants 
\begin{equation}
p_1=\lam_A,\; p_2=x_1 x_5,\; p_3=s^4 ~.
\end{equation}
Note that we could have added to $p_2$ a multiple of $(p_3)^2$, so that it could be replaced by $\lam_B$ for instance.

As an alternative for the ansatze derivation  one could have used maximal minors to compute the 2-minors of $G_{cf}$, obtain an invariant 3-form, and attempt to factorize it into a wedge product of three differentials of the invariants.  

\presub {\bf Homogeneous solution}. By definition, $I_0$, the homogeneous solution of (\ref{PDEs}) must be a homogenous solution of the constant-free set, and hence must be a function of the constant free invariants \be
I_0 = I_0(p_1,p_2,p_3) ~.
\ee
Substituting this ansatz into the remaining equations we obtain the following equation system for $I_0$
\begin{eqnarray}
(d-3)I_0 - 2p_1 \frac{\partial}{\partial p_1} I_0 &=& 0 \\
 I_0 + 2 p_2 \frac{\partial}{\partial p_2} I_0 + p_3\frac{\partial}{\partial p_3} I_0 &=& 0 ~.
\end{eqnarray}
Its general solution is \be
 I_0 = g(\phi)\, \frac{\lambda_A^{\frac{d-3}{2}}}{s^4}~.\label{homsol}
\ee
where $g(\phi)$ is an arbitrary function of the $G$-invariant, which we choose to set to unity.

\presub {\bf Line integral representation}. In general, SFI allows to replace the integral at any point in parameter space by the integral at a conveniently chosen point on the same $G$-orbit plus a line integral over simpler diagrams.

Here we shall employ property (\ref{base}) and take our base point to have $m_2=m_3=0$. At this point the integration over the bubble (the loop containing propagators 2 and 3) is immediate, and it remains to evaluate a variant of the bubble diagram. This value appears in the literature -- see (\ref{massless23}).

Next we should integrate the simpler source terms from $m_2=m_3=0$ to the point of interest. This can be done over any contour, and in particular we can take a line or a piecewise linear contour. Performing variation of the constants \be
 I(x) = c(x)\, I_0(x)
\ee
 and substituting into equations $E^2, E^3$ (\ref{def:E}) of the SFI equation system (\ref{PDEs}) to solve for $\del^2 c,\, \del^3 c$ we obtain the following line integral representation \bea
  I(x) &=&\lam_A^{\frac{d-3}{2}}\, \left\{ \frac{\left. I \right|_0 }{(x_1)^{d-3}}  +  2  \int_0^x \[ \frac{dx'_2}{\lam_A^{\frac{d-1}{2}}} \(x_3 \del^3(O_2-O_1)I-s^1_A \del^2 (O_3-O_1) I \) + \(2 \leftrightarrow 3 \) \]_{x'} \right\} \non 
  \label{line_int}
  \eea
where the evaluation point $0$ denotes $x_2=x_3=0$ and the integration path is taken with fixed $x_1, x_4, x_5$.  For the integrand to be well defined we assume that $\lam_A(x) \ge 0$ so that the $\lam_A$ factor in the denominator starts at $\lam_A=(x_1)^2 \ge 0$ and never crosses a zero along the path.  
In the special case where $x_3=0$ the expression simplifies to \be
I(x) =  (x_1-x_2)^{d-3} \[  \frac{\left. I \right|_{x_2=0}}{(x_1)^{d-3}}  +  \int_{x_2=0}^x  \frac{dx'_2}{(x_1-x'_2)^{d-2}} \left. \del^2 (O_3-O_1) I) \right|_{x'_2 }  \] ~.
\label{line_int_zero_x3}
\ee
Similarly here we assume that $x_2 \le x_1$.


\section{Sector evaluation}
\label{sec:sectors}

In this section we consider certain special sectors in parameter space where the sources appearing in the general line integral (\ref{line_int_zero_x3}) are known and we are able to evaluate the integral to obtain more explicit expressions.

As a 4-scale sector we consider the sector with $x_3=0$. This is the only sector where sunrise diagrams appearing in the source terms have at least one massless propagator and hence their value is known in terms of Appell functions.

Proceeding to 3-scale sectors we consider setting in addition also $x_4$ or $x_1$ to zero. The remaining case $x_2=x_3=0$ is known in terms of Appell functions (\ref{massless23}), 
 and we do not consider it.

\subsection{Massless $m_3, m_4$}
\label{masslessm3m5}
\begin{figure}[t]
\begin{center}
\includegraphics[scale=0.25]{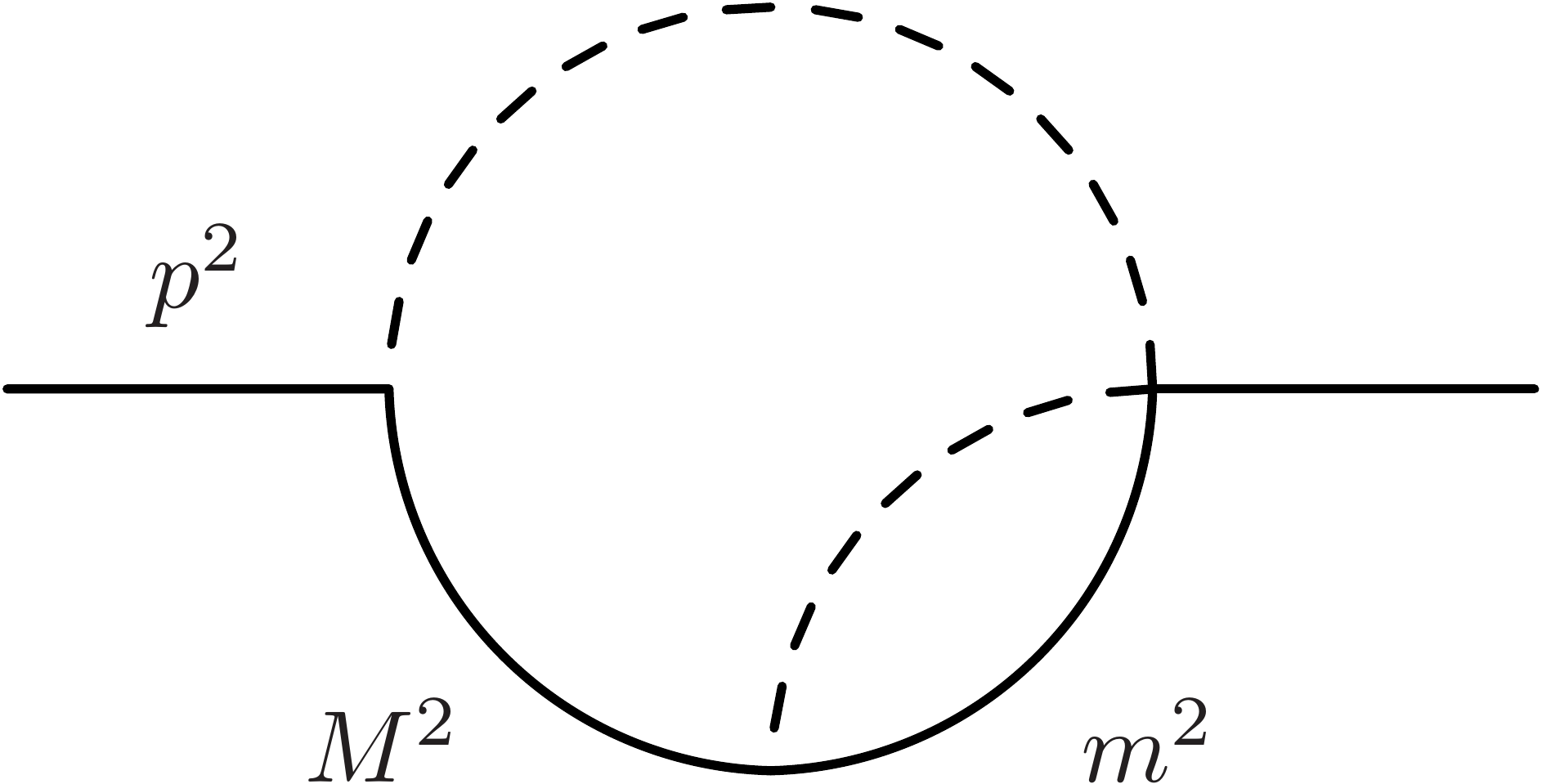}
\caption{The diagram for sector 1, namely  $I(M^2,m^2,0,0,p^2)$. The dashed lines denote massless propagators.}
\label{psm1m2p}
\end{center}
\end{figure}

Let us calculate the integral at the sector with $x_3=x_4=0$ parameterized by $B_1=(M^2,m^2,0,0,p^2)$ (see Fig.~\ref{psm1m2p}), starting from $A_1=(M^2,0,0,0,p^2)$ and proceeding along the path
$\gamma_1=(M^2,m^2 t,0,0,p^2), 0\leq t\leq 1$. 
Note that $A_1$, $B_1$ and $\gamma_1$ are on the orbit with
\begin{equation}\label{phi124}
\phi=\frac{M^2-p^2}{\sqrt{M^2p^2}}=const~.
\end{equation}

The line integral (\ref{line_int_zero_x3}) is given by
\begin{eqnarray}
I(M^2,m^2,0,0,p^2)&=& (M^2-m^2)^{d-3}\Big[\int_0^1 \frac{\partial^2 O_3I(M^2,m^2t,p^2)-\partial^2 O_1I(m^2t,p^2)}{(M^2-m^2t)^{d-2}}m^2 dt\nonumber\\
& &+\frac{I(M^2,0,0,0,p^2)}{(M^2)^{d-3}}\Big]~.
\end{eqnarray}
The sources in this case are (see (\ref{g}),(\ref{jtad}) and (\ref{jpm}) for definitions)
\begin{eqnarray}
\partial^2 O_1I(m^2t,p^2)&=&i^{1-d}\pi^{d/2}G(1,1)\tilde{J}_{bubble}(2-d/2,2;0,m^2t,p^2)\nonumber \\
&=&i^{2-2d}\pi^{d}(p^2)^{d-4}\Big[a_1(d)\, {_{2}F_{1}}\left(4-d,5-\frac{3d}{2};3-\frac{d}{2}\Big| \frac{m^2}{p^2}t\right)\nonumber\\
& &+a_2(d)\left(-\frac{m^2}{p^2}t\right)^{d/2-2} {_{2}F_{1}}\left(2-\frac{d}{2},3-d,\frac{d}{2}-1\Big| \frac{m^2}{p^2}t\right)\Big]\\
\partial^2 O_3I(M^2,m^2t,p^2)&=&\tilde{J}_{bubble}(1,1;0,M^2,p^2)J_{tad}(2;m^2t)\nonumber \\
&=&i^{2-2d}\pi^{d} (p^2)^{d/2-2}(-m^2t)^{d/2-2}\Big[a_{3a}(d)\Big(1-\frac{M^2}{p^2}\Big)^{d-3}\nonumber\\
&&+a_{3b}(d)\Big(-\frac{M^2}{p^2}\Big)^{d/2-1}{_{2}F_{1}}\left(1,2-\frac{d}{2},\frac{d}{2}\Big|\frac{M^2}{p^2}\right)\Big]
\end{eqnarray}
where
\begin{eqnarray}
a_1(d)&=&G(1,1)G(2-d/2,2)\\
a_2(d)&=& a_{3a}(d) =G(1,1)\, \Gamma(2-d/2)\\
a_{3b}(d)&=&\Gamma(2-d/2)\Gamma(1-d/2)~.
\end{eqnarray}
The starting point can be calculated through the Mellin-Barnes method and is given by 
\begin{eqnarray}
I(M^2,0,0,0,p^2)
& =& -  i^{2-2d}\pi^{d}(p^2)^{d-4} 
\Big[\tilde{a}_1(d) {_{2}F_{1}} \Big( 1, 5-\frac{3d}{2},3-\frac{d}{2} \Big| \frac{M^2}{p^2}\Big) \nonumber \\
& & +\tilde{b}_1(d) \Big(-\frac{M^2}{p^2}\Big)^{d-3}  {_{2}F_{1}} \Big( 1,  2-\frac{d}{2}, \frac{d}{2}\Big| \frac{M^2}{p^2}\Big) \Big]
\end{eqnarray}
where $\tilde{a}_{1a}(d)$ is given in (\ref{a1tilde}) and $\tilde{b}_1(d)$ is given in (\ref{b1tilde}).

We see that the only non-trivial integrals we have to evaluate are of the form $\int_0^1 \frac{t^\beta{_{2}F_{1}}\left(a,b,c | x t\right)}{\left(1-y t\right)^{\alpha}}dt$, and they can be computed by expressing ${_{2}F_{1}}$ by its power series definition and exchanging the order of integration and summation
\begin{eqnarray}
& &\int_0^1 \frac{t^\beta\,{_{2}F_{1}}\left(a,b,c | x t\right)}{\left(1-y t\right)^{\alpha}}dt \nonumber \\
&&=\sum_{k=0}^\infty \frac{(a)_k(b)_k}{(c)_k}\frac{x^k}{k!}\int_0^1 \frac{t^{\beta+k}}{(1-yt)^\alpha}dt \nonumber \\
&&=\sum_{k=0}^\infty \frac{(a)_k(b)_k}{(c)_k}\frac{x^k}{k!}\frac{1}{1+\beta+k}{_{2}F_{1}}(\alpha,1+\beta+k,2+\beta+k|y)\nonumber\\
&& =\Gamma(1-\alpha)\Big[\frac{\Gamma(1+\beta)}{\Gamma(2+\beta-\alpha)}y^{-1-\beta}\sum_{k=0}^\infty \frac{(a)_k(b)_k(1+\beta)_k}{(c)_k(2-\alpha+\beta)_k}\frac{(x/y)^k}{k!} \nonumber\\
&& -\frac{(1-y)^{1-\alpha}}{\Gamma(2-\alpha)}\sum_{k=0}^\infty \sum_{n=0}^\infty \frac{(2-\alpha+\beta)_{k+n}(a)_k(b)_k(1)_n}{(2-\alpha+\beta)_k(c)_k(2-\alpha)_n}\frac{x^k}{k!}\frac{(1-y)^n}{n!}   \Big]
\end{eqnarray} 
where the fourth line is gotten by applying the hypergeometric function identity ${_{2}F_{1}}(a,b,c,z)= \frac{\Gamma(c - a - b) \Gamma(c)}{\Gamma(c - a) \Gamma(c - b)} {_{2}F_{1}}(a, b, a + b + 1 - c, 1 - z) + \frac{\Gamma(a + b - c) \Gamma(c)}{\Gamma(a) \Gamma(b)} (1 - z)^{c - a - b} {_{2}F_{1}}(c - a, c - b, c + 1 - a - b, 1 - z)$.
The first single sum can be recognized as the hypergeometric function ${_{3}F_{2}}$.  Once we plug in the values for $a,b,c,d,\alpha,\beta$, it becomes ${_{2}F_{1}}$ while the second sum simplifies to the Appell $F_1$ or Appell $F_2$ function. 

Our final result is
\begin{eqnarray}\label{result_sector1}
&&I(M^2,m^2,0,0,p^2)= i^{2-2d}\pi^d \left[ I_1(M,m,p)+I_2(M,m,p)+I_3(M,m,p)\right]
\end{eqnarray}
where
\begin{eqnarray}
I_1(M,m,p)&=&-\tilde{a}_{1a}(d)(p^2)^{d-4} F_1\left( 5-3d/2 , 3-d,1 ,3-d/2 \,\Big| \frac{m^2}{p^2},\frac{M^2}{p^2}\right)\nonumber\\
&&-\tilde{a}_{1b}(d)(-p^2 m^2)^{\frac{d}{2}-2} \frac{m^2}{M^2}F_2\left(2-\frac{d}{2},3-d,1,\frac{d}{2}-1,4-d \Big| \frac{m^2}{p^2}, \, 1-\frac{m^2}{M^2}\right)\nonumber\\
&&-\tilde{a}_{1c}(d)(m^2M^2)^{\frac{d}{2}-2}\frac{m^2}{p^2}\, \left(1-\frac{m^2}{M^2}\right)^{d-3}\nonumber \\
&&\times{_{2}F_{1}}\left(\frac{d}{2}-1,d-2,\frac{d}{2} \Big| \frac{m^2}{M^2}\right){_{2}F_{1}}\left(1,2-\frac{d}{2},\frac{d}{2} \Big| \frac{M^2}{p^2}\right)\nonumber \\ \\
I_2(M,m,p)&=&\left(1-\frac{m^2}{M^2}\right)^{d-3}\Big[\tilde{a}_{2a}(d)(-p^2m^2)^{\frac{d}{2}-2}\frac{m^2}{M^2}\left(1-\frac{M^2}{p^2}\right)^{d-3} {_{2}F_{1}}\left(\frac{d}{2}-1,d-2,\frac{d}{2} \Big| \frac{m^2}{M^2}\right)\nonumber \\
&&-\tilde{a}_{2b}(d) (p^2)^{d-4} \Big(-\frac{M^2}{p^2}\Big)^{d-3}  {_{2}F_{1}} \Big( 1,  2-\frac{d}{2}, \frac{d}{2}\Big| \frac{M^2}{p^2}\Big) \Big]\\
I_3(M,m,p)&=& \tilde{a}_{3}(d)(-p^2M^2)^{\frac{d}{2}-2} \left(1-\frac{m^2}{M^2}\right)^{d-3} \left(1-\frac{M^2}{p^2}\right)^{d-3}
\end{eqnarray}

with
\begin{eqnarray}
\tilde{a}_{1a}(d)&=&-2\frac{\Gamma^3(d/2-1)\Gamma(3-d)}{(d-4)\Gamma(3d/2-4)} \label{a1tilde}\\
\tilde{a}_{1b}(d)&=&\frac{2\pi\, \Gamma(3-d)}{(d-3)\, \tan(d\pi/2)}\\
\tilde{a}_{1c}(d)&=&(-1)^{d+1}\Gamma^2(1-d/2)\\
\tilde{a}_{2a}(d)&=&\frac{2\pi^2}{\sin^2(d\pi/2)\Gamma(d-1)}\\
\tilde{a}_{2b}(d)&=&\tilde{b}_1(d)=-2\pi \frac{\Gamma(2-d)}{ \sin\left(d\,\pi/2\right)}\\
\tilde{a}_{3}(d)&=&\frac{\pi \Gamma(3-d)\Gamma^2(d/2-1)}{\sin(d\pi/2)\Gamma(d-2)}.
\end{eqnarray}
We have compared this result with a Mellin-Barnes computation to find full numerical agreement at arbitrary points in parameter space.

\subsection{Massless $m_3, m_1$}

\begin{figure}[t]
\begin{center}
\includegraphics[scale=0.25]{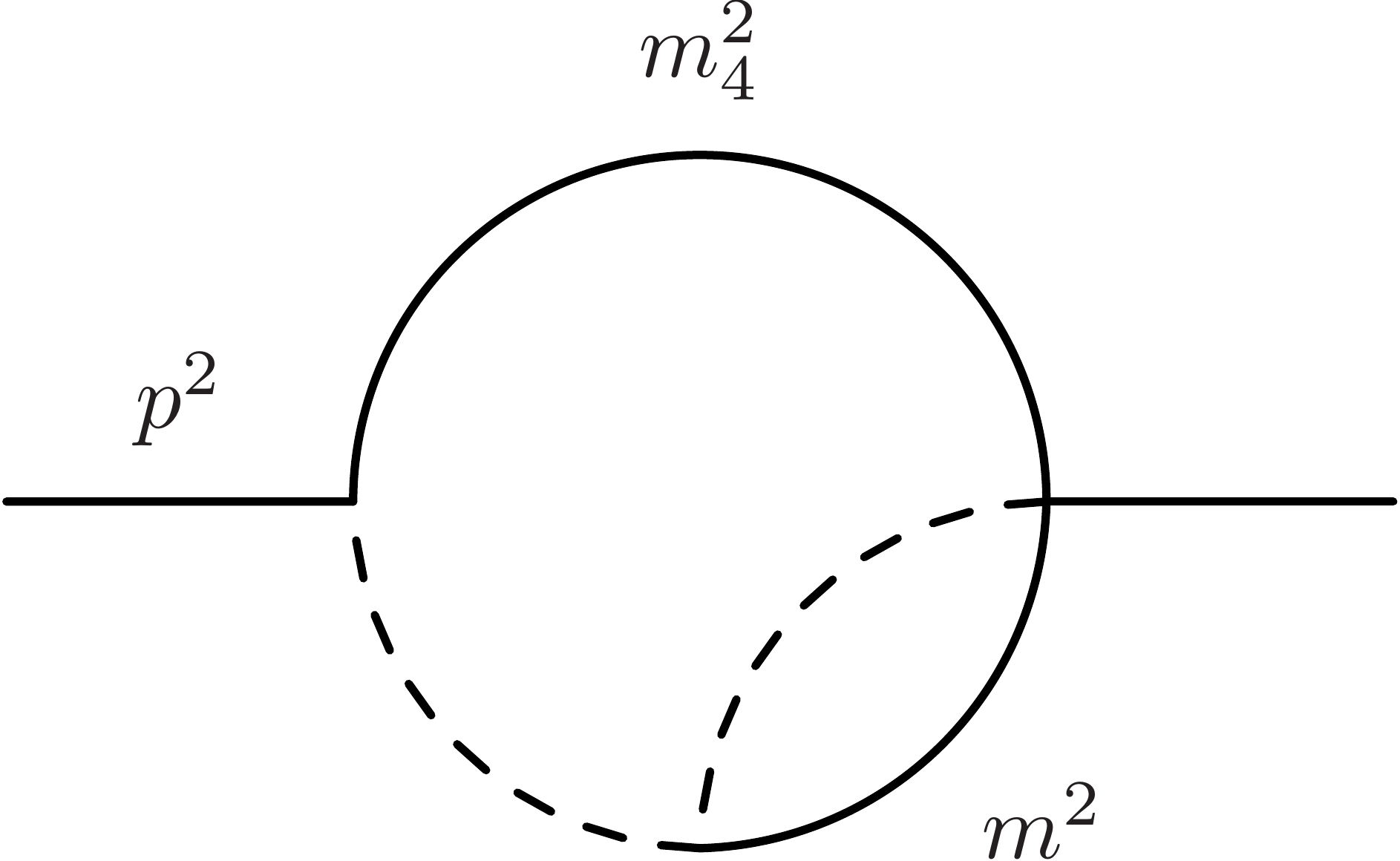}
\caption{The diagram for sector 2, namely $I(0,m^2,0,m_4^2,p^2)$.}
\label{psm2m5p}
\end{center}
\end{figure}

Now we shall calculate the propagator seagull at the sector with $x_1=x_3=0$ parameterized by $B_2=(0,m^2,0,m_4^2,p^2)$, see Fig.~\ref{psm2m5p}. Our starting point will be $A_2=(0,0,0,m_4^2,p^2)$ and the path will be $\gamma_2=(0,m^2\,t,0,m_4^2,p^2)$ with $0\leq t\leq 1$. Note that this trajectory is on $\phi = \infty$. Here our line integral (\ref{line_int_zero_x3}) is given by
\begin{eqnarray}
I(0,m^2,0,m_4^2,p^2)&=&\int_0^1 \frac{\partial^2 O_1I(m^2t,m_4^2,p^2)-\partial^2 O_3I(m^2t,m_4^2,p^2)}{t^{d-2}} dt\nonumber \\
&&+(m^2)^{d-3}\lim_{M^2\to0}\frac{I(M^2,0,0,m_4^2,p^2)}{(M^2)^{d-3}}\Big]~.
\end{eqnarray}

Since $I_0$ vanishes at the starting point $A_2$, its contribution to the integral diverges and we regulate it by setting an IR cut-off $x_1=M^2$ and taking the limit $M^2 \to 0$.

The sources in this case are given by (see (\ref{jtad}), (\ref{jpm1}), (\ref{jpmM}) for definitions)
\begin{eqnarray}
\partial^2 O_1I(m^2t,m_4^2,p^2)&=&J_{sunrise}(1,2,1;0,m^2 t,m_4^2,p^2)\nonumber \\
&=&i^{2-2d}\pi^{d}(-m_4^2)^{d-4}\nonumber\\
&&\times \Big[b_1(d)F_4\left(3-\frac{d}{2},4-d,\frac{d}{2},3-\frac{d}{2}|\frac{p^2}{m_4^2},\frac{m^2 t}{m_4^2}\right)\nonumber \\
&+&b_2(d)\left(\frac{m^2}{m_4^2}\right)^{d/2-2}t^{d/2-2}F_4\left(1,2-\frac{d}{2},\frac{d}{2},\frac{d}{2}-1|\frac{p^2}{m_4^2},\frac{m^2 t}{m_4^2}\right)\Big]  \label{j12mpm5} \\
\partial^2 O_3I(m^2t,m_4^2,p^2)&=&J_{bubble}(1,1;0,m_4^2,p^2)J_{tad}(2;m^2t)\nonumber \\
&=&i^{2-2d}\pi^{d}b_3(d)(-m_4^2)^{d/2-2}(-m^2)^{d/2-2}t^{d/2-2}{_{2}F_{1}}\left(1,2-\frac{d}{2},\frac{d}{2}|\frac{p^2}{m_4^2}\right)\nonumber\\
\end{eqnarray}
where 
\begin{eqnarray}
b_1(d)&=&\frac{2}{d-2}\pi \csc\left(\frac{\pi d }{2}\right)\Gamma(4-d)\\
b_2(d)&=&\frac{2}{d-2}\Gamma ^2\left(2-\frac{d}{2}\right)\\
b_3(d)&=& \frac{d-2}{2} \Gamma ^2\left(1-\frac{d}{2}\right).
\end{eqnarray}
The integral over $\partial^2 O_3I(t)$ is trivial. For the integral over $\partial^2 O_1I(t)$ we find that we need to perform integrals of the type

\begin{eqnarray*}
& &\int_0^1 \frac{F_4(a,b,a-b+1,d|x,y\,t)}{t^\alpha}dt=\int_0^1t^{-\alpha}\sum_{k=0}^{\infty}\sum_{n=0}^{\infty} \frac{(a)_{k+n}(b)_{k+n}}{k!\,n!\,(a-b+1)_k(d)_n}x^k(y\,t)^n\,dt\nonumber\\
&&=\frac{1}{1-\alpha}\sum_{k=0}^{\infty}\sum_{n=0}^{\infty}\frac{(a)_{k+n}(b)_{k+n}(1-\alpha)_n}{k!\,n!\,(a-b+1)_k(d)_n(2-\alpha)_n}x^ky^n\nonumber \\
&&=\frac{1}{1-\alpha}\sum_{n=0}^{\infty}\frac{(a)_n(b)_n(1-\alpha)_n}{n!(d)_n(2-\alpha)_n}y^n{_{2}F_{1}}(a+n,b+n,a-b+1|x)\nonumber\\
&&=\frac{1}{1-\alpha}\sum_{n=0}^{\infty}\frac{(a)_n(b)_n(1-\alpha)_n}{n!(d)_n(2-\alpha)_n}y^n(1\pm \sqrt{x})^{-2(a+n)}{_{2}F_{1}}\left(a+n,a-b+\frac{1}{2} ,2a-2b+1|\frac{\pm 4\sqrt{x}}{(1\pm \sqrt{x})^2}\right)\nonumber\\
&&=\frac{1}{1-\alpha}(1\pm \sqrt{x})^{-2a}\sum_{k=0}^\infty\sum_{n=0}^{\infty}\frac{(a)_{k+n}(a-b+\frac{1}{2})_k(b)_n(1-\alpha)_n}{k!\,n!\,(2a-2b+1)_k(d)_n(2-\alpha)_n}\left(\frac{\pm 4\sqrt{x}}{(1\pm \sqrt{x})^2}\right)^k \left(\frac{ y}{(1\pm \sqrt{x})^2}\right)^n \nonumber\\
\end{eqnarray*}
where we used the quadratic transformation, ${_{2}F_{1}}(a,b,a-b+1|z)=\\
\left(1\pm \sqrt{z}\right)^{-2a} {_{2}F_{1}}\left(a,a-b+\frac{1}{2},2a-2b+1\Big|\pm \frac{4\sqrt{z}}{(1\pm \sqrt{z})^2}\right)$, to reach the fourth line. Again, once we plug in the values for $a,b,c,d,\alpha$ we find that this double sum simplifies to an Appell $F_2$ function.

The starting point is taken as the limit $\lim_{M^2\to0}\frac{I(M^2,0,0,m_4^2,p^2)}{(M^2)^{d-3}}$ where $I(M^2,0,0,m_4^2,p^2)$ can be calculated via Mellin-Barnes 
\cite{Bauberger:1994by} (for a possible simplification from $F_2$ and $F_4$ functions to $F_1$ and ${_{2}F_1}$ see \cite{Davydychev:2000na} footnote 16):
\begin{eqnarray}\label{massless23}
I(M^2,0,0,m_4^2,p^2)&=&i^{2-2d}\pi^d(-1)^{d-4}\, \tilde{b}_1(d) \Big[m_4^{d-2}(m_4\pm p)^{d-6} \nonumber \\
&& \times F_2\left(3-\frac{d}{2},\frac{d}{2}-\frac{1}{2}, 1, d-1, 3-\frac{d}{2} \Big|  \frac{\pm 4p\,m_4}{(m_4 \pm p)^2},\, \frac{ M^2}{(m_4 \pm p)^2}\right) \nonumber\\
& &-\frac{(M^2)^{d-3}}{m_4^2}F_4\left(1,\frac{d}{2},\frac{d}{2},\frac{d}{2}\Big|\frac{p^2}{m_4^2},\frac{M^2}{m_4^2}\right)\Big]~.
\end{eqnarray}
$\tilde{b}_1(d)$ is given in (\ref{b1tilde}). Here and in the following $\pm$ means that either sign can be chosen consistently throughout the expression and we are denoting $\sqrt{p^2}\equiv p$. The limit
\begin{eqnarray}
\lim_{M^2\to0}\frac{I(M^2,0,0,m_4^2,p^2)}{(M^2)^{d-3}}=- i^{2-2d}\pi^d(-1)^{d-4} \tilde{b}_1(d) \frac{1}{m_4^2-p^2}~.
\end{eqnarray}

In this way we arrive at the complete result
\begin{eqnarray}\label{result_sector2}
I(0,m^2,0,m_4^2,p^2)=i^{2-2d}\pi^d(-1)^{d-4}\Big[ I_1(m,m_4,p)+I_2(m,m_4,p)\nonumber \\+I_3(m,m_4,p)+I_4(m,m_4,p)\Big]
\end{eqnarray}
where
\begin{eqnarray}
I_1(m,m_4,p)&=& \tilde{b}_1(d)(m_4)^{d-2}\left(m_4 \pm p \right)^{d-6}\nonumber \\
&&\times F_2\left(3-\frac{d}{2}, \frac{d}{2}-\frac{1}{2}, 3-d,d-1,3-\frac{d}{2}\Big| \frac{\pm 4p\,m_4}{(m_4 \pm p)^2},\, \frac{ m^2}{(m_4 \pm p)^2}\right)\nonumber \\ \\
I_2(m,m_4,p)&=&\tilde{b}_2(d)\frac{(m\;m_4)^{d-2}}{(m(m_4 \pm p))^2}\nonumber \\
&&\times F_2\left(1, \frac{d}{2}-\frac{1}{2}, 1-\frac{d}{2},d-1,\frac{d}{2}-1\Big| \frac{\pm 4p\,m_4}{(m_4 \pm p)^2},\, \frac{ m^2}{(m_4 \pm p)^2}\right)\nonumber \\ \\
I_3(m,m_4,p)&=&-\tilde{b}_2(d)(m\; m_4)^{d-4}{_{2}F_{1}}\left(1,2-\frac{d}{2},\frac{d}{2}\Big|\frac{p^2}{m_4^2}\right)\\
I_4(m,m_4,p)&=&-\tilde{b}_1(d)\frac{(m^2)^{d-3}}{m_4^2-p^2}
\end{eqnarray}
with
\begin{eqnarray}
\tilde{b}_1(d)&=&-2\pi \frac{\Gamma(2-d)}{ \sin\left(d\,\pi/2\right)} \label{b1tilde}\\
\tilde{b}_2(d)&=&- \Gamma ^2\left(1-\frac{d}{2}\right).
\end{eqnarray}
Also in this case we have compared this result with a Mellin-Barnes computation to find full numerical agreement at arbitrary points in parameter space.

\subsection{Massless $m_3$}

\begin{figure}[t]
\begin{center}
\includegraphics[scale=0.25]{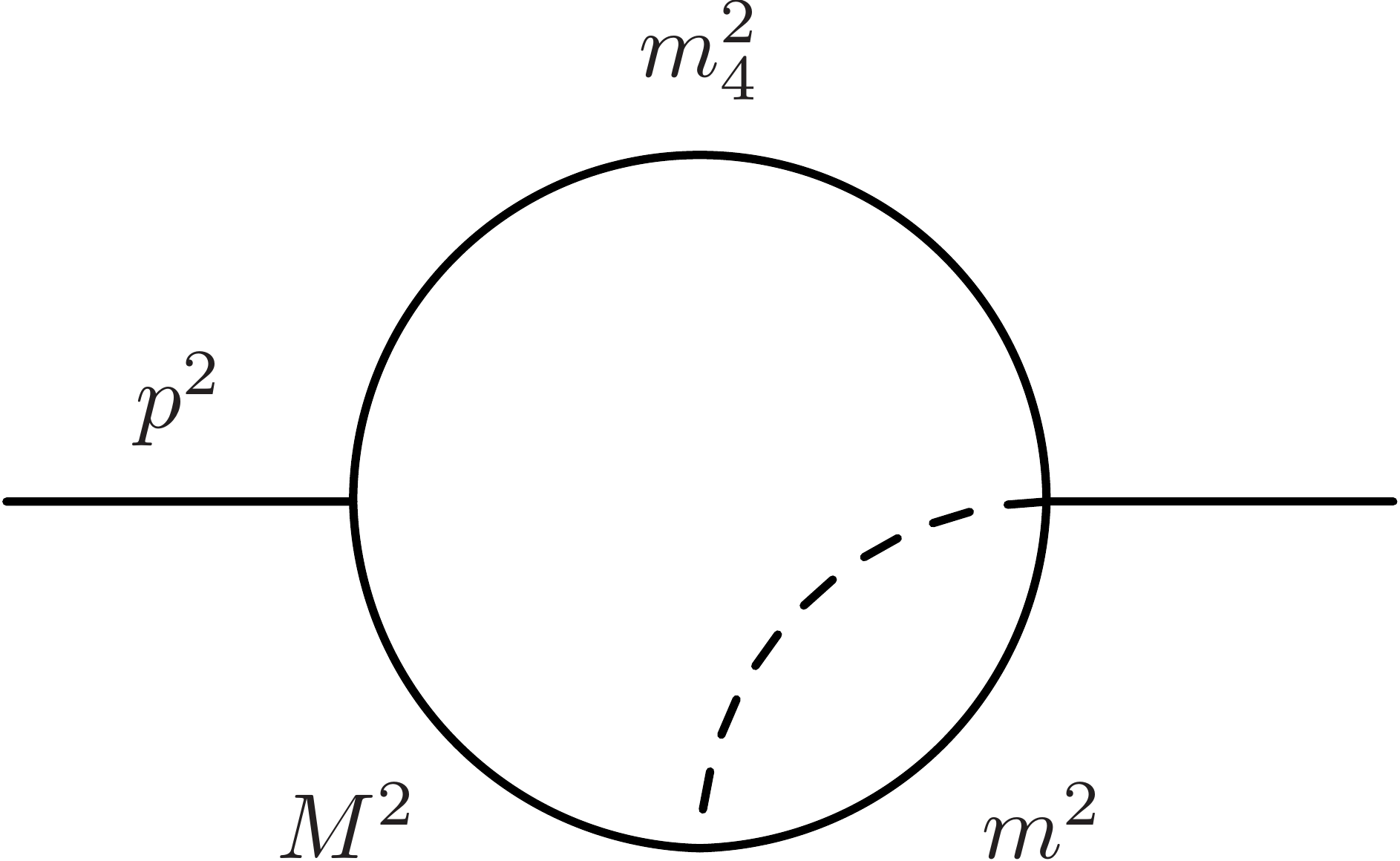}
\caption{The diagram for sector 3, namely $I(M^2,m^2,0,m_4^2,p^2)$.}
\label{psm1m2m5p}
\end{center}
\end{figure}

Here we give a 4 scale result. The starting point is $A_3=(M^2,0,0,m_4^2,p^2)$ and the final point lies on the $x_3=0$ subspace parameterized by $B_3=(M^2,m^2,0,m_4^2,p^2)$, see Fig.~\ref{psm1m2m5p}. The path is given by $\gamma_3=(M^2,m^2 t,0,m_4^2,p^2)$ with $0\leq t \leq 1$.
The integral to perform is 
\begin{eqnarray}
I(M^2,m^2,0,m_4^2,p^2)&=&\left(1-\frac{m^2}{M^2}\right)^{d-3}\nonumber \\
&&\times \Big[\frac{m^2}{M^2}\int_0^1 \frac{\partial^2 O_3I(M^2,m^2 t,m_4^2,p^2)-\partial^2 O_1I(m^2 t,0,m_4^2,p^2)}{\left(1-\frac{m^2}{M^2}t\right)^{d-2}}dt\nonumber\\
&&+I(M^2,0,0,m_4^2,p^2)\Big].
\end{eqnarray}
Here
\begin{eqnarray}
\partial^2 O_1I(m^2 t,0,m_4^2,p^2)&=&J_{sunrise}(1,2,1;0,m^2 t,m_4^2,p^2)
\end{eqnarray}
defined in (\ref{jpmM}) and given explicitly in Eq. (\ref{j12mpm5}) and
\begin{eqnarray}
\partial^2 O_3I(M^2,m^2 t,m_4^2,p^2)&=& J_{bubble}(1,1;M^2,m_4^2,p^2)J_{tad}(2;m^2 t)
\end{eqnarray}
with $J_{tad}$ and $J_{bubble}$ defined in (\ref{jtad}) and (\ref{jbubble}) and given explicitly by
\begin{eqnarray}
 &&J_{bubble}(1,1;M^2,m_4^2,p^2)\nonumber \\
 &&=i^{1-d}\pi^{d/2}(-m_4^2)^{\frac{d}{2}-2}\Big[-\frac{\pi}{\sin(\pi\,d/2)\Gamma(d/2)} F_4\left(1,2-\frac{d}{2},\frac{d}{2},2-\frac{d}{2} \Big| \frac{p^2}{m_4^2},\frac{M^2}{m_4^2}\right)\nonumber \\
 &&+\Gamma \left(1-\frac{d}{2} \right) \left(\frac{M^2}{m_4^2}\right) ^{d/2-1}F_4\left(1,\frac{d}{2},\frac{d}{2},\frac{d}{2} \Big|  \frac{p^2}{m_4^2},\frac{M^2}{m_4^2}\right)\Big] 
\end{eqnarray}
and
\begin{equation}
J_{tad}(2;m^2 t)=i^{1-d}\pi^{d/2} \Gamma\left(2-\frac{d}{2}\right) (-m^2)^{d/2-2}\,t^{d/2-2}~.
\end{equation}
Integrating $\partial^2 O_3I(t)$ amounts to the integral
\begin{eqnarray}
\int_0^1 \frac{t^{d/2-2}}{\left(1-\frac{m^2}{M^2}t\right)^{d-2}}dt=\frac{2}{d-2}{_{2}F_{1}}\left(\frac{d}{2}-1,d-2,\frac{d}{2}\Big|\frac{m^2}{M^2}\right)~.
\end{eqnarray}

The integral over $\partial^2 O_1I(t)$ is of the form
\begin{eqnarray*}
& &\int_0^1 \frac{t^\beta F_4(a,b,c,d|x,y\,t)}{ (1-z\,t)^\alpha}dt=
 \sum_{k=0}^{\infty}\sum_{n=0}^{\infty} \frac{(a)_{k+n}(b)_{k+n}}{k!n!(c)_k(d)_n}x^k y^n \int_0^1 \frac{t^{\beta+n}}{(1-zt)^{\alpha}}dt \nonumber \\
= &&\frac{1}{1+\beta} \sum_{k=0}^{\infty}\sum_{n=0}^{\infty} \frac{(a)_{k+n}(b)_{k+n}(1+\beta)_n}{k!\,n!\,(a-b+1)_k(d)_n(2+\beta)_n}x^k y^n {_{2}F_{1}}(\alpha,1+\beta+n,2+\beta+n|z) \nonumber \\
=&&\frac{1}{1+\beta}\sum_{n=0}^\infty \frac{(a)_n(b)_n(1+\beta)_n}{n!\,(d)_n(2+\beta)_n} y^n {_{2}F_{1}}(a+n,b+n,a-b+1|x){_{2}F_{1}}(\alpha,1+\beta+n,2+\beta+n|z)\nonumber\\
=&&\frac{1}{1+\beta}\frac{\Gamma(2+\beta)\Gamma(1-\alpha)}{\Gamma(2-\alpha+\beta)}z^{-1-\beta}(1\pm \sqrt{x})^{-2a} \nonumber \\
&&\times \sum_{l=0}^{\infty}\sum_{n=0}^{\infty}\frac{(a)_{l+n}(a-b+\frac{1}{2})_l(b)_n(1+\beta)_n}{l!\,n!\,(2a-2b+1)_l(d)_n(2-\alpha+\beta)_n}\left(\frac{\pm 4\sqrt{x}}{(1\pm \sqrt{x})^2}\right)^l\left(\frac{y}{z(1\pm \sqrt{x})^2}\right)^n\nonumber \\
&&+\frac{1}{\alpha-1}(1-z)^{1-\alpha}(1\pm \sqrt{x})^{-2a} \sum_{l=0}^{\infty}\sum_{m=0}^{\infty}\sum_{n=0}^{\infty}\frac{(a)_{l+n}(2-\alpha+\beta)_{m+n}(a-b+\frac{1}{2})_l(1)_m(b)_n}{l!\,m!\,n!\,(2-\alpha)_m(2a-2b+1)_l(d)_n(2-\alpha+\beta)_n}\nonumber \\
&&\times \left(\frac{\pm 4\sqrt{x}}{(1\pm \sqrt{x})^2}\right)^l  (1-z)^m \left(\frac{y}{(1\pm \sqrt{x})^2}\right)^n~.
\end{eqnarray*}
In the fourth line we used both the quadratic transformation ${_{2}F_{1}}(a,b,a-b+1|z)=\\
\left(1\pm \sqrt{z}\right)^{-2a} {_{2}F_{1}}\left(a,a-b+\frac{1}{2},2a-2b+1\Big|\pm \frac{4\sqrt{z}}{(1\pm \sqrt{z})^2}\right)$ and the identity ${_{2}F_{1}}(a,b,c,z)= \frac{\Gamma(c - a - b) \Gamma(c)}{\Gamma(c - a) \Gamma(c - b)} {_{2}F_{1}}(a, b, a + b + 1 - c, 1 - z) + \frac{\Gamma(a + b - c) \Gamma(c)}{\Gamma(a) \Gamma(b)} (1 - z)^{c - a - b} {_{2}F_{1}}(c - a, c - b, c + 1 - a - b, 1 - z)$. Here, the double sum simplifies to ${_{2}F_{1}}$ or an Appell $F_2$ function and the triple sum to a Lauricella $F_K$ function, see (\ref{lauricella}) for the definition of $F_K$.

The starting point in this sector, $I(M^2,0,0,m_4^2,p^2)$ is given by (\ref{massless23}).

Finally our result is
\begin{eqnarray}\label{result_sector3}
I(M^2,m^2,0,m_4^2,p^2)&=&i^{2-2d}\pi^{d}\big[I_1(M,m,m_4,p)+I_2(M,m,m_4,p)+I_3(M,m,m_4,p) \big]\nonumber \\ 
\end{eqnarray}
where
\begin{eqnarray}
&&I_1(M,m,m_4,p)=(-1)^{d-4}\tilde{c}_1(d)\,m_4^{d-2}\left(m_4 \pm p \right)^{d-6} 
\frac{m^2}{M^2}\nonumber \\
&&\times F_K\left(3-\frac{d}{2},4-d,1,\frac{d}{2}-\frac{1}{2},d-1,4-d,3-\frac{d}{2}\Big|\frac{\pm 4p\,m_4}{(m_4 \pm p)^2},1-\frac{m^2}{M^2},\frac{m^2}{(m_4 \pm p)^2} \right)
\nonumber \\
&&+(-1)^{d-4}\tilde{c}_{2}(d)\frac{(m\,m_4)^{d-2}}{(M(m_4 \pm p))^2}\nonumber \\ 
&&\times F_K\left(1,2-\frac{d}{2},\frac{d}{2}-\frac{1}{2},1,d-1,4-d,\frac{d}{2}-1\Big|\frac{\pm 4p\,m_4}{(m_4 \pm p)^2},1-\frac{m^2}{M^2},\frac{m^2}{(m_4 \pm p)^2}  \right)\nonumber \\ \\
&&I_2(M,m,m_4,p)=(-1)^{d-4}\tilde{c}_{1}(d)\,m_4^{d-2}M^{d-4} \left((m_4 \pm p)^2-M^2\right)^{-1} \left(1-\frac{m^2}{M^2}\right)^{d-3}\nonumber \\
&&\times {_{2}F_{1}}\left(1,\frac{d}{2}-\frac{1}{2},d-1 \Big|\frac{\pm 4m_4\,p}{(m_4 \pm p)^2-M^2}\right) \\
&&I_3(M,m,m_4,p)=i^{d-1}\pi^{-d/2}\tilde{c}_3(d)\frac{m^2}{M^2}\left(1-\frac{m^2}{M^2}\right)^{d-3}\,{_{2}F_{1}}\left(\frac{d}{2}-1,d-2,\frac{d}{2}\Big|\frac{m^2}{M^2}\right)\nonumber \\ 
&&\times J_{bubble}(1,1;M^2,m_4^2,p^2)
\end{eqnarray}

with
\begin{eqnarray}
\tilde{c}_1(d)&=&\tilde{b}_1(d)=-2\pi \frac{\Gamma(2-d)}{ \sin\left(d\,\pi/2\right)}\\
\tilde{c}_{2}(d)&=&\frac{\Gamma(1-d/2)\Gamma(2-d/2)}{d-3}\\
\tilde{c}_3(d)&=&\frac{2}{d-2}.
\end{eqnarray}
We have confirmed this result by comparing numerically with a Mellin-Barnes computation. It would be interesting to test this result by restricting it to $m_4=0$ and compare with the result of Section \ref{masslessm3m5}.

\subsection{$\epsilon$ expansion with massless $m_3$} \label{ep_expansion}
In this section we compute analytically the $\epsilon$ expansion for the propagator seagull with four arbitrary scales, namely $m_1^2,m_2^2,m_4^2$ and $p^2$ up to order $\epsilon^0$. We will expand the line integral result (\ref{line_int_zero_x3}) around $d=4-2\epsilon$. We will need the $\epsilon$ expansion of the source terms and of the starting point which will be taken from the literature.

We rewrite (\ref{line_int_zero_x3}) in the following way
\begin{eqnarray}
 I(x_1,x_2,0,x_4,p^2)&=&(x_1-x_2)^{d-3}\Big[ \frac{I(x_1,0,0,x_4,p^2)}{x_1^{d-3}}\nonumber \\&&+x_2\int_0^1\frac{\Bub(x_1,x_4,p^2)\partial \Tad(x_2 t)-\partial \Sun(x_2 t,x_4,p^2)}{(x_1-tx_2)^{d-2}}dt \Big].\non \label{line_integral_ep}
\end{eqnarray}
Since we are considering here the $\epsilon$ expansion of the source terms we will denote them by a different notation from that used in the general $d$ calculations. Here $\Bub(x_1,x_4,p^2)$ is the bubble diagram with masses squared $x_1,x_4$, $\Tad(x_2)$ is the tadpole diagram with mass squared $x_2$ and $\Sun(x_2,x_4,p^2)$ is the sunrise diagram with one massless propagator and two massive propagators $x_2,x_4$ (see appendix A for the definitions of bubble, tadpole and sunrise). The integral from 0 to $x_2$ over $x_2'$ is replaced by an integral from 0 to 1 by scaling $x_2'=tx_2$.
We have identified $\partial^2O_3I=\Bub(x_1,x_4,p^2)\partial_{x_2} \Tad(x_2)$ and $\partial^2O_1I=\partial_{x_2} \Sun(x_2,x_4,p^2)$. We will also use the renormalization scale $Q^2=4\pi e^{-\gamma}\mu^2$ to define logarithms of dimensionful quantities.  

Taken from \cite{Martin:2003qz,Martin:2005qm}, the $\epsilon$ expansions of the source terms are given by
\begin{equation}
\partial_x \Tad(x)=\frac{1}{\epsilon}-\partial_x A(x)- \partial_x A_\epsilon (x)\epsilon +\sc{O}(\epsilon^2)
\end{equation}
\begin{equation}
\Bub(x,y,p^2)=\frac{1}{\epsilon}+B(x,y,p^2)+ B_\epsilon (x,y,p^2)\epsilon+\sc{O}(\epsilon^2)
\end{equation}
\begin{equation}
\partial_x \Sun(x,y,p^2)=\frac{1}{2\epsilon^2}-\left(\frac{A(x)}{x}+\frac{1}{2}\right)\frac{1}{\epsilon}-\partial S(x,y,p^2)+\frac{A(x)-A_\epsilon(x)}{x}+\sc{O}(\epsilon)
\end{equation}

where 
\begin{equation}
A(x)=x\left(\ln\left(\frac{x}{Q^2}\right)-1\right), \quad \partial_x A(x)=\ln\left(\frac{x}{Q^2}\right)
\end{equation}
\begin{equation}
A_\epsilon(x)=x\left(-1-\frac{\zeta(2)}{2}+\ln\left(\frac{x}{Q^2}\right)-\frac{1}{2}\ln^2\left(\frac{x}{Q^2}\right)\right), \quad\partial_x A_\epsilon (x)=-\frac{1}{2}\left(\zeta(2)+\ln^2\left(\frac{x}{Q^2}\right)\right).
\end{equation}
The functions $B(x_1,x_4,p^2)$, $B_{\epsilon}(x_1,x_4,p^2)$ are defined in appendix \ref{appendix_epsilon_expansion} in (\ref{bfunction},\ref{befunction}), and $\partial S(x,y,p^2)$ is defined below in equation (\ref{ds_function}).

The $\epsilon$ expansion of the starting point is \cite{{Scharf:1993ds},{Martin:2005qm}}
\begin{equation}
I(x_1,0,0,x_4,p^2)=\frac{1}{2\epsilon^2}+\left(\frac{1}{2}+B(x_1,x_4,p^2)\right)\frac{1}{\epsilon}+I(x_1,x_4,p^2)+\sc{O}(\epsilon) \label{stp}.
\end{equation}
$I(x_1,x_4,p^2)$ is defined in appendix \ref{appendix_epsilon_expansion} in (\ref{ifunction}). The form of $B(x_1,x_4,p^2)$, $B_{\epsilon}(x_1,x_4,p^2)$ and $I(x_1,x_4,p^2)$ does not affect the integration.

Expansion of the other parts of (\ref{line_integral_ep}) in $\epsilon$ gives
\begin{align}
\frac{(x_1-x_2)^{d-3}}{x_1^{d-3}}= \left(1-\frac{x_2}{x_1}\right)-  \, 2\left(1-\frac{x_2}{x_1}\right)\ln \left(1-\frac{x_2}{x_1}\right)\epsilon+2\left(1-\frac{x_2}{x_1}\right)\ln^2\left(1-\frac{x_2}{x_1}\right)\epsilon^2+\dots \nonumber
\end{align}
and
\begin{align*}
&\frac{(x_1-x_2)^{d-3}}{{(x_1-tx_2)}^{d-2}}= (x_1-x_2)\Big[ \frac{1}{(x_1-t x_2)^2}+ \frac{2\left(\ln(x_1-t x_2)-\ln(x_1-x_2)\right)}{(x_1-t x_2)^2} \epsilon\\
&+  \frac{2\left(\ln(x_1-t x_2)-\ln(x_1-x_2)\right)^2}{(x_1-t x_2)^2}\epsilon^2 +\dots\nonumber\Big].
 \nonumber
\end{align*}
Now we are ready to compute the $\epsilon$ expansion of the propagator seagull up to order $\epsilon^0$.

\presub{\bf $\epsilon^{-2}$ term}.
After collecting terms of order $\epsilon^{-2}$ we find 
\begin{eqnarray}
&&\Big[ \left(1-\frac{x_2}{x_1}\right)+x_2(x_1-x_2)\int_0^1 \frac{1}{(x_1-t x_2)^2}dt  \Big]\frac{1}{2\epsilon^2}=\frac{1}{2\epsilon^2} \label{ep-2_term}.
\end{eqnarray}

\presub{\bf $\epsilon^{-1}$ term}.
Collecting terms of order $\epsilon^{-1}$ we get after some simplification
\begin{eqnarray}
&&\frac{1}{2\epsilon}(x_1-x_2)\Big[-\frac{2}{x_1}\ln(1-x_2/x_1)+\frac{2B(x_1,x_4,p^2)}{x_1}+\frac{1}{x_1}\nonumber \\
&&+(-x_2-2x_2\ln(x_1-x_2)+2x_2 B(x_1,x_4,p^2)\int_0^1 \frac{1}{(x_1-tx_2)^2}dt\nonumber \\
&&+2x_2\int_0^1 \frac{\ln(x_1-t x_2)}{(x_1-t x_2)^2}dt\Big]=\frac{1}{2\epsilon}\left(1+2B(x_1,x_4,p^2)\right).\label{ep-1_term}
\end{eqnarray}

\presub{\bf $\epsilon^{0}$ term}.
Collecting terms and performing some integrations we are left with
\begin{eqnarray}
&&\frac{x_2}{x_1}+\left(1-\frac{x_2}{x_1}\right)I(x_1,x_4,p^2)+\frac{x_2}{x_1}B_\epsilon(x_1,x_4,p^2)\nonumber \\
&&+\Big[\frac{2x_2}{x_1}-\frac{x_2}{x_1}\ln\left(\frac{x_2}{Q^2}\right)-\left(1-\frac{x_2}{x_1}\right)\ln\left(1-\frac{x_2}{x_1}\right)\Big] B(x_1,x_4,p^2)\nonumber \\
&&+P(x_1,x_2,x_4,p^2) . \label{ep0_term}
\end{eqnarray}
The final term is the more involved integral 
\begin{equation}
P(x_1,x_2,x_4,p^2)\equiv x_2(x_1-x_2) \int_0^1 \frac{\partial S(tx_2,x_4,p^2)}{(x_1-t x_2)^2}dt
\end{equation}
where
\begin{eqnarray}
\partial S(x,y,p^2)&=&\Li_2\left(t(y,p^2,x)\right)+\Li_2\left(r(y,p^2,x)\right)-\left(1-\frac{y}{p^2}\right)\ln\left(t(y,x,p^2)\right)\ln\left(r(y,x,p^2)\right)\nonumber \\
&&+\frac{\sqrt{\lambda(x,y,p^2)}}{2p^2}\left(\ln \left(t(y,x,p^2)\right)-\ln \left(r(y,x,p^2)\right)\right)+\frac{y-x-p^2}{2p^2}\ln\left(\frac{x}{y}\right)\nonumber \\
&&-\ln\left(\frac{x}{Q^2}\right)\left(\ln\left(\frac{y}{Q^2}\right)-2\right)+\frac{1}{2}\left(\ln \left( \frac{y}{Q^2}\right)-1\right)^2  \label{ds_function}
\end{eqnarray}
and
\begin{align}
t(x,y,z)\equiv t_{xyz}&=\frac{x+y-z+\sqrt{\lambda(x,y,z)}}{2x},\\
r(x,y,z)\equiv r_{xyz}&=\frac{x+y-z-\sqrt{\lambda(x,y,z)}}{2x}
\end{align}
are the roots of the equation $x u^2- (x+y-z)u+y=0$. 

We successfully performed the integral to obtain
\begin{align}
&P(x_1,x_2,x_4,p^2)\equiv x_2(x_1-x_2) \int_0^1 \frac{\partial S(tx_2,x_4,p^2)}{(x_1-t x_2)^2}dt=\Li_2(t_{4p2})+\Li_2(r_{4p2})\nonumber\\
&+\frac{\sqrt{\lambda_{24p}}}{2p^2}(\ln(t_{42p})-\ln(r_{42p}))-\frac{x_2}{2p^2}\ln(x_2/x_4)-\frac{x_2}{x_1}\left(\ln(x_4/Q^2)-2\right)\ln(x_2/Q^2)\nonumber\\
&+\frac{x_2}{2x_1}\left(1-\frac{x_4}{p^2}\right)(\ln^2(t_{42p})+\ln^2(r_{42p})-\ln^2(x_2/x_4)-\ln(x_2/x_4))\nonumber\\
&+\frac{x_2}{2x_1}\left(\ln(x_4/Q^2)-1\right)^2+(x_1-x_2)P_1(x_1,x_2,x_4,p^2)\label{p_function}
\end{align}
where $P_1(x_1,x_2,x_4,p^2)$ is given by
\begin{small}
\begin{align}
&P_1(x_1,x_2,x_4,p^2)=-\frac{1}{x_1}\Big[\Li_2(t_{4p0})+\Li_2(r_{4p0})+\frac{\sqrt{\lambda_{04p}}}{2p^2}(\ln(t_{40p})-\ln(r_{40p}))\nonumber \\
&+\left(1-\frac{x_4}{p^2}\right)(\ln(1-x_2/x_1)\left(1/2+\ln(x_2/x_4)\right)+\Li_2(x_2/x_1))+\left(\ln(x_4/Q^2)-2\right)\ln(1-x_2/x_1)\Big]\nonumber\\
&-\frac{1}{2p^2}\left(\Li_2(x_2/x_1)+\ln(1-x_2/x_1)(1+\ln(x_2/x_4))\right)\nonumber \\
&+\frac{1}{\sqrt{\lambda_{14p}}}\Big[\Li_2\left(\frac{t_{4p2}-r_{4p1}}{1-r_{4p1}}\right)-\Li_2\left(\frac{t_{4p2}-t_{4p1}}{1-t_{4p1}}\right)-\Li_2\left(\frac{t_{4p0}-r_{4p1}}{1-r_{4p1}}\right)+\Li_2\left(\frac{t_{4p0}-t_{4p1}}{1-t_{4p1}}\right)\nonumber  \\
&+\ln(1-t_{4p1})\left(\ln(t_{4p2}-t_{4p1})-\ln(t_{4p0}-t_{4p1})\right)-\ln(1-r_{4p1})\left(\ln(t_{4p2}-r_{4p1})-\ln(t_{4p0}-r_{4p1})\right)\nonumber  \\
&+\Li_2\left(\frac{r_{4p2}-r_{4p1}}{1-r_{4p1}}\right)-\Li_2\left(\frac{r_{4p2}-t_{4p1}}{1-t_{4p1}}\right)-\Li_2\left(\frac{r_{4p0}-r_{4p1}}{1-r_{4p1}}\right)+\Li_2\left(\frac{r_{4p0}-t_{4p1}}{1-t_{4p1}}\right)\nonumber  \\
&+\ln(1-t_{4p1})\left(\ln(r_{4p2}-t_{4p1})-\ln(r_{4p0}-t_{4p1})\right)-\ln(1-r_{4p1})\left(\ln(r_{4p2}-r_{4p1})-\ln(r_{4p0}-r_{4p1})\right)\Big]\nonumber  \\
&+\frac{p^2-x_4}{p^2 \sqrt{\lambda_{14p}}}\Big[\left(1-\frac{1}{t_{41p}}\right)\Big(\ln(t_{42p})\ln(1-t_{42p}/t_{41p})+\Li_2(t_{42p}/t_{41p})\nonumber  \\
&-\ln(t_{40p})\ln(1-t_{40p}/t_{41p})-\Li_2(t_{40p}/t_{41p})+\ln(r_{42p})\ln(1-r_{42p}/t_{41p})+\Li_2(r_{42p}/t_{41p})\nonumber \\
&-\ln(r_{40p})\ln(1-r_{40p}/t_{41p})-\Li_2(r_{40p}/t_{41p})\Big)-\left(1-\frac{1}{r_{41p}}\right)\Big(\ln(t_{42p})\ln(1-t_{42p}/r_{41p})\nonumber\\
&+\Li_2(t_{42p}/r_{41p})-\ln(t_{40p})\ln(1-t_{40p}/r_{41p})-\Li_2(t_{40p}/r_{41p})\nonumber  \\
&+\ln(r_{42p})\ln(1-r_{42p}/r_{41p})+\Li_2(r_{42p}/r_{41p})-\ln(r_{40p})\ln(1-r_{40p}/r_{41p})-\Li_2(r_{40p}/r_{41p})\Big)\Big]\nonumber\\
&+\frac{1}{2p^2}\Big[-\Li_2(1-t_{42p})+\Li_2(1-t_{40p})-\Li_2(1-r_{42p})+\Li_2(1-r_{40p})\nonumber\\
&+\frac{x_4-p}{x_1}(\ln(t_{42p})-\ln(t_{40p})+\ln(r_{42p})-\ln(r_{40p}))\nonumber \\
&+\left(1-\frac{1}{t_{41p}}\right)(\ln(t_{42p}-t_{41p})-\ln(t_{40p}-t_{41p})+\ln(r_{42p}-t_{41p})-\ln(r_{40p}-t_{41p}))\nonumber \\
&+\left(1-\frac{1}{r_{41p}}\right)(\ln(t_{42p}-r_{41p})-\ln(t_{40p}-r_{41p})+\ln(r_{42p}-r_{41p})-\ln(r_{40p}-r_{41p}))\Big]\nonumber \\
&+\frac{x_1-x_4-p^2}{2p^2\sqrt{\lambda_{14p}}}\Big[ \ln(t_{42p})(\ln(1-t_{42p}/t_{41p})-\ln(1-t_{42p}/r_{41p}))\nonumber\\
&+\Li_2(t_{42p}/t_{41p})-\Li_2(t_{42p}/r_{41p})-\ln(t_{40p})(\ln(1-t_{40p}/t_{41p})-\ln(1-t_{40p}/r_{41p}))\nonumber\\
&-\Li_2(t_{40p}/t_{41p})+\Li_2(t_{40p}/r_{41p})+\ln(r_{42p})(\ln(1-r_{42p}/t_{41p})-\ln(1-r_{42p}/r_{41p}))\nonumber\\
&+\Li_2(r_{42p}/t_{41p})-\Li_2(r_{42p}/r_{41p})-\ln(r_{40p})(\ln(1-r_{40p}/t_{41p})-\ln(1-r_{40p}/r_{41p}))\nonumber\\
&-\Li_2(r_{40p}/t_{41p})+\Li_2(r_{40p}/r_{41p})\Big]\label{p1_function}.
\end{align}
\end{small}
In (\ref{p1_function})  we used the notation $t_{40p}=t(x_4,0,p), \lambda_{04p}=\lambda(0,x_4,p)$ etc.

We successfully compared the result (\ref{p_function}) with a numerical integration at randomly chosen parameters. Note that when $x_2\to 0$, Eq.\ (\ref{ep0_term}) reduces to $I(x_1,x_4,p^2)$ which is the $\epsilon^0$ term of $I(x_1,0,0,x_4,p^2)$ as it should be (see (\ref{stp})). We believe this result (\ref{ep0_term},\ref{p_function},\ref{p1_function}) for the $\epsilon^0$ term of $I(x_1,x_2,0,x_4,p^2)$ to be new.

Note that \cite{Martin:2003qz} gives the $\epsilon^0$ term for a different 4-scale sector which in our notation is $I(x_1,x_2,x_3,0,p^2)$.

\section{Singular locus}
\label{al}

For any diagram, there could be special hyper-surfaces in parameter space, called the singularity locus, where the SFI equation system becomes degenerate \cite{locus}.  On this locus the equations can be combined in such a way that the differential part cancels. In \cite{minors} it was shown how to systematically find this singular locus using maximal minors.

The singularity locus of the SFI equation system (\ref{PDEs}) is defined by setting \be
 0= S \equiv \lam_A
\label{sing1}
 \ee
 where the singular factor $S$ was obtained in (\ref{def:S}) through the computation of 4-minors. 
 
\presub {\bf $\lam_A$ locus}. At $\lam_A=0$ there exists a group stabilizer, namely, a row vector that is a left null vector of $Tx$. It is found to be \be
 Stb_{\lam_A} = (0,\, -x_3,\, s^1_A,\, 0 ) \parallel (0,\,  s^1_A,\, -x_2,\, 0 ) (\textrm{mod } \lam_A)
 \ee 
where the two forms are parallel on the $\lam_A=0$ locus, namely mod $\lam_A$. Multiplying the equations system by $Stb_{\lam_A}$ the differential part of the equation cancels and we obtain an algebraic equation for $I$ which is solved to yield \be
I(x) \rvert_{\lambda_A} = \frac{1}{(d-3)\, s^2} \( x_3 \partial^3 (O_2- O_1)I - s^1_A \partial^2 (O_3- O_1)I \) \equiv \( 2 \leftrightarrow 3 \).
\label{I_lamA1}
\ee

Another equivalent expression can be written involving square roots. We use $s^1_A = \pm \sqrt{x_2 x_3} \text{ mod } \lam_A$ and two other identities gotten by permutations. We must be careful about signs and recall that the $\lam_A=0$ cone is divided into three parts, where in each part exactly one $s$ variable is negative, see e.g. \cite{diameter} Fig.~3. In this way we get \be
I(x)\rvert_{\lambda_A} = \frac{1}{d-3} \Big( \eps_2\, \sqrt{\frac{x_2}{x_1}} \partial^2 (O_3- O_1)I+\eps_3\, \sqrt{\frac{x_3}{x_1}} \partial^3 (O_2- O_1)I \Big)
\label{I_lamA2}
\ee
where the signs $\eps_2,\, \eps_3$ are given by \be
(\eps_2,\, \eps_3)= \begin{cases}
(+,+) & \text{for } s^1\le 0 \\ 
(+,-) &   \text{for } s^2 \le 0 \\
(-,+) & \text{for } s^3 \le 0
\end{cases} .
\label{signs}
\ee
This expression is manifestly symmetric  under $2 \leftrightarrow 3$ exchange.

We checked the expressions (\ref{I_lamA1},\ref{I_lamA2}) with \texttt{FIRE} \cite{Smirnov:2014hma} at the point $x_1=x_3=m^2, x_2=0$ and general $x_4, x_5$.

\presub{\bf Co-dimension 2 loci}. Let us consider co-dimension 2 hyper-surfaces in parameter space where the equation system turns algebraic.  This happens when the invariant 1-form (\ref{def:inv}) 
 degenerates, namely  $Inv=0$, and therefore so do the 4-minors. This locus is composed of two components which we now turn to consider.  

Component 2a is defined by $x_1=0$ and $x_4=x_5$. The stabilizer is \be
Stb_{2a} = \begin{pmatrix} 
	\half (x_3-x_2) \, , & -x_3 \, ,	& x_2 \, , 	& \half (x_2-x_3)
 \end{pmatrix} 
\ee
 and the algebraic solution is
 \be
 I(x)\rvert_{2a} = \frac{2}{x_2-x_3} \( x_2 \del^2 (O_3-O_1)I - (2 \leftrightarrow 3) \) - \del^4 O_1 I ~.
 \label{I_2a}
 \ee
We have checked this result with the computer package \texttt{FIRE} \cite{Smirnov:2014hma}.
 
Component 2b is defined by $x_5 \equiv p^2 = 0$ and $x_1=x_4$.
Here the stabilizer is  simply \be
Stb_{2b} = \begin{pmatrix} 
	0 \, , &    0 \, ,	& 0	\, , & 1
 \end{pmatrix} 
\ee
 and accordingly the algebraic solution is
 \be
 I(x)\rvert_{2b} =  \del^4 O_1 I ~.
  \label{I_2b}
 \ee
This equation means that setting $p^2=0$ is equivalent to erasing the external legs and further setting $x_1=x_4$ leads us to the diameter diagram with a squared propagator.

We note that in general, co-dimension 2 algebraic loci occur also at the intersection of algebraic locus components, for example in the vacuum seagull.

\section{Summary and Discussion}
\label{sec:summary}

In this paper we have analyzed the propagator seagull diagram for arbitrary values of its parameters: the masses and $p^2$, and for general spacetime dimension $d$, obtaining the following results: \bi

\item The SFI equation system of differential equations was determined (\ref{PDEs}). 
The associated group $G \subset GL(3,\IR)$ was found to be 4d and non-simple, and is shown in (\ref{trans}). 

\item The $G$-orbits were shown to be defined by a single invariant, $\phi$, given by (\ref{phi_value}) (or functions thereof) 
 and interpreted geometrically as an angle in the on-shell triangle associated with the trivalent vertex $B$. 

\item A general reduction was given in (\ref{line_int}) in terms of a base point where $m_2=m_3=0$ together with a line integral over simpler diagrams. 

\item Explicit expressions in terms of special functions, including the hypergeometric, Appell and Lauricella functions, were obtained for a pair of 3-scale sectors in (\ref{result_sector1},\ref{result_sector2}) and for a 4-scale sector in (\ref{result_sector3}). 
The three expressions were confirmed against an independent computation using the Mellin-Barnes transform of massive propagators. 

\item The $\epsilon$ expansion up to order $\epsilon^0$ for the 4-scale sector was calculated in subsection (\ref{ep_expansion}) and is given in (\ref{ep-2_term}), (\ref{ep-1_term}) and (\ref{ep0_term},\ref{p_function},\ref{p1_function}).

\item The singular locus was determined, both at co-dimension~1 and at co-dimension~2. In co-dimension 1 the locus has a single component given by $\lam_A=0$,  and the integral is represented by simpler diagrams in (\ref{I_lamA1}) or (\ref{I_lamA2}). 
 The co-dimension 2 locus was found to be composed of two components. The first is at $x_1=p^2-x_4=0$ where the solution is given by (\ref{I_2a}), and the second is at  $p^2=x_1-x_4=0$ and the solution is (\ref{I_2b}).  Certain confirmations for these results were achieved through \texttt{FIRE} \cite{Smirnov:2014hma}.

\ei

The results are interesting both with respect to the diagram and with respect to the SFI method. Regarding the diagram,  sector results withstood several tests and we believe that they are novel and naturally interesting. Regarding the SFI method, this case is our first example where we find a group invariant, and it provides valuable experience with performing the reduction to a line integral and with analysis of the singular locus.

\presub {\bf Discussion}. It is interesting to compare the propagator seagull (PS) with the vacuum seagull (VS). Thanks to the adopted numbering convention for propagators, the SFI equation system for PS is identical with 4 of the 5 equations of VS. The fifth equation is forbidden in PS anyway in the sense of the last comment in section \ref{sec:eq_system}. The singular locus of PS is $S_{PS} = \lam_A$ which is one of the components of the singular locus of VS, namely $S_{VS} = x_1\, \lam_A\, \lam_B$.

\presub {\bf Open questions}. 

{\bf Value at base point}. As a base manifold we chose to set $m_2=m_3=0$. At this value the integral reduces to a sort of bubble integral which is given by (\ref{massless23}). It would be interesting to analyze this integral through SFI.

{\bf Sectors expressions: simplification and analysis}. The sector expressions include several special functions. It would be interesting to analyze the expressions, for example, to study their analytic structure. Another interesting question is whether a simplification of the expressions might be possible.

{\bf Allowing for numerator}. The propagator seagull has an ISP which can appear as a numerator. We leave it for future work to analyze the integral with this numerator.

{\bf Relation with Landau equations}. The singular locus of the SFI equation system appears to be closely related to the Landau equations. It would be very interesting to establish this relation.   

\subsection*{Acknowledgments}
 
We would like to thank Philipp Burda, Subhajit Mazumdar and Amit Schiller for discussions. RS is very grateful to Philipp Burda for participation in the early stages of this work.
 
This research was  supported by the ``Quantum Universe'' I-CORE program of the Israeli Planning and Budgeting Committee.

\appendix

\section{Collection of function definitions}
\label{sec:defin}
Here we collect the functions used in the definition of $\partial^2 O_1I,\partial^2 O_3I$ in Sec.~\ref{sec:sectors}. As shown in (\ref{sources}) the possible topologies for the source diagrams are that of ``tadpole" $\times$ ``bubble'' and ``sunrise''. Where  ''tadpole"$\equiv \raisebox{-7pt}{\includegraphics[scale=0.2]{tadpole_w_ex}} $, ``bubble"$\equiv \raisebox{-12pt}{\includegraphics[scale=0.3]{1loop_bubble}}$ and ``sunrise"$\equiv \raisebox{-12pt}{\includegraphics[scale=0.3]{2loop_sunset}}$\,.

The diagram with tadpole topology with general powers on the loop propagator is given by
\begin{equation}
J_{tad}(n;m^2)=i^{1-d}\pi^{d/2}\frac{\Gamma(n-d/2)}{\Gamma(n)}(-m^2)^{d/2-n}\label{jtad}~.
\end{equation}
The bubble topology with general masses $m_1, m_2$ and powers $n_1, n_2$ on the propagators is given by (see e.g. \cite{Boos:1990rg})
\begin{eqnarray}\label{jbubble}
J_{bubble}(n_1,n_2;m_1^2,m_2^2,p^2)&=&\pi^{d/2}i^{1-d}(-m_2^2)^{d/2-n_1-n_2}\Big[\frac{\Gamma(d/2-n_1)\Gamma(n_1+n_2-d/2)}{\Gamma(d/2)\Gamma(n_2)}\nonumber \\
& & F_4\left(n_1,n_1+n_2-\frac{d}{2},\frac{d}{2},n_1+1-\frac{d}{2} \Big| \frac{p^2}{m_2^2},\frac{m_1^2}{m_2^2}\right)\nonumber \\
&&+\left(\frac{m_1^2}{m_2^2}\right)^{d/2-n_1}\frac{\Gamma(n_1-d/2)}{\Gamma(n_1)}F_4\left(n_2,\frac{d}{2},\frac{d}{2},\frac{d}{2}+1-n_1 \Big| \frac{p^2}{m_2^2},\frac{m_1^2}{m_2^2}\right)\Big].\nonumber \\
\end{eqnarray}
When one of the propagators is massless this expression simplifies to
\begin{eqnarray}
J_{bubble}(n_1,n_2;0,m^2,p^2)&=&i^{1-d}\pi^{d/2}(-m^2)^{d/2-n_1-n_2}\frac{\Gamma(d/2-n_1)\Gamma(n_1+n_2-d/2)}{\Gamma(d/2)\Gamma(n_2)}\nonumber \\
& &{_{2}F_{1}}\left(n_1,n_2+n_2-\frac{d}{2},\frac{d}{2}\Big|\frac{p^2}{m^2}\right)\label{jpm1}
\end{eqnarray}
and by analytic continuation is given also by the following expression, which we find easier to use in Sec.~\ref{masslessm3m5}
\begin{eqnarray}
& &\tilde{J}_{bubble}(n_1,n_2;0,m^2,p^2)=i^{1-d}\pi^{d/2}\Big[ G(n_1,n_2){_{2}F_{1}}\Big(n_1+n_2-\frac{d}{2},n_1+n_2+1-d,n_2+1-\frac{d}{2}\Big| \frac{m^2}{p^2}\Big)\nonumber\\
& &+\Big(-\frac{m^2}{p^2}\Big)^{d/2-n_2}\frac{\Gamma(n_2-\frac{d}{2})}{\Gamma(n_2)} {_{2}F_{1}}\Big(n_1,n_1+1-\frac{d}{2},\frac{d}{2}+1-n_2\Big| \frac{m^2}{p^2}\Big)
\Big](p^2)^{d/2-n_1-n_2}\label{jpm}
\end{eqnarray}
where
\begin{equation}
G(n_1,n_2)=\frac{\Gamma(n_1+n_2-d/2)\Gamma(d/2-n_1)\Gamma(d/2-n_2)}{\Gamma(n_1)\Gamma(n_2)\Gamma(d-n_1-n_2)}.\label{g}
\end{equation}
The sunrise diagram with one massless propagator and general powers $n_1,n_2,n_3$ on the propagators is given by (see \cite{Jegerlehner:2003py})
\begin{eqnarray}
&&J_{sunrise}(n_3,n_2,n_1;0,m^2,M^2,p^2)=
i^{2-2d}\pi^d\,(-M^2)^{d-n_3-n_2-n_1}\Big[\left(\frac{\Gamma(d/2-n_3)}{\Gamma(n_3)\Gamma(n_1)\Gamma(n_2)\Gamma(d/2)}\right)\nonumber \\
&& \Gamma\left(\frac{d}{2}-n_2\right)\Gamma(n_1+n_2+n_3-d)\Gamma\left(n_3+n_3-\frac{d}{2}\right)\nonumber \\
&&\times F_4\left(n_2+n_3-\frac{d}{2},n_1+n_2+n_3-d,\frac{d}{2},1+n_2-\frac{d}{2} \Big| \frac{p^2}{M^2},\frac{m^2}{M^2}\right)+\nonumber \\
&&\left(\frac{m^2}{M^2}\right)^{d/2-n_2}\Gamma\left(n_2-\frac{d}{2}\right)\Gamma(n_3)\Gamma\left(n_1+n_3-\frac{d}{2}\right)\nonumber \\
&& \times F_4\left(n_3+n_1-\frac{d}{2},n_3,\frac{d}{2},1+\frac{d}{2}-n_2 \Big| \frac{p^2}{m^2},\frac{m^2}{M^2}\right)\Big]\label{jpmM}
\end{eqnarray}
where $F_4$ is the fourth Appell hypergeometric function.

The Lauricella $F_K$ function is defined by
\begin{eqnarray}\label{lauricella}
F_K(a,b,c,d,e,f|x,y,z)=\sum_{l=0}^\infty \sum_{m=0}^\infty \sum_{n=0}^\infty \frac{(a)_{l+n}(b)_{m+n}(c)_l(d)_m}{(d)_l(e)_m(f)_n}\frac{x^l}{l!} \frac{y^m}{m!} \frac{z^n}{n!}
\end{eqnarray}
where $(a)_n=a(a+1)\cdot \dots \cdot (a+n-1)$ is the Pochhammer symbol. 

\section{Functions appearing in the $\epsilon$ expansion}\label{appendix_epsilon_expansion}

The functions $B(x,y,p^2)$ and $B_\epsilon(x,y,p^2)$ are the $\epsilon^0$ and $\epsilon^1$ terms in the $\epsilon$ expansion of the bubble and are given by \cite{Scharf:1993ds} in Eq.\ (83)
\begin{align}
B(x_1,x_4,p^2)=&-\frac{1}{2}\Big[\ln\left(\frac{x_1}{Q^2}\right)+\ln\left(\frac{x_4}{Q^2}\right)-4+\frac{x_1/x_4-1}{p^2/x_4}\ln\left(\frac{x_1}{x_4}\right)\nonumber \\
&-\frac{r_{41p}-t_{41p}}{p^2/x_4}(\ln(r_{41p})-\ln(t_{41p}))\Big]\label{bfunction}
\end{align}
\begin{align}
&B_\epsilon(x_1,x_4,p^2)=\frac{1}{2}\Big\{\zeta(2)+8+\frac{1}{4}\left(\ln\left(\frac{x_1}{Q^2}\right)+\ln\left(\frac{x_4}{Q^2}\right)\right)^2+\frac{1}{4}\ln^2\left(\frac{x_1}{x_4}\right)\nonumber \\
&+\left(\ln\left(\frac{x_1}{Q^2}\right)+\ln\left(\frac{x_4}{Q^2}\right)\right)\left(-2+\frac{x_1/x_4-1}{2p^2/x_4}\ln\left(\frac{x_1}{x_4}\right)-\frac{r_{41p}-t_{41p}}{2p^2/x_4}(\ln(r_{41p})-\ln(t_{41p}))\right)\nonumber \\
&-2\frac{x_1/x_4-1}{p^2/x_4}\ln\left(\frac{x_1}{x_4}\right)+\frac{r_{41p}-t_{41p}}{p^2/x_4}\Big[2(\ln(r_{41p})-\ln(t_{41p}))\nonumber\\
&+\ln\left(\frac{1-r_{41p}}{t_{41p}-r_{41p}}\right)\ln\left(\frac{-r_{41p}(1-t_{41p})}{t_{41p}-r_{41p}}\right)
-\ln\left(\frac{1-t_{41p}}{r_{41p}-t_{41p}}\right)\ln\left(\frac{-t_{41p}(1-r_{41p})}{r_{41p}-t_{41p}}\right)\nonumber\\
&+\Li_2\left(\frac{-r_{41p}(1-t_{41p})}{t_{41p}-r_{41p}}\right)-\Li_2\left(\frac{-t_{41p}(1-r_{41p})}{r_{41p}-t_{41p}}\right)-\Li_2\left(\frac{1-t_{41p}}{r_{41p}-t_{41p}}\right)+\Li_2\left(\frac{1-r_{41p}}{t_{41p}-r_{41p}}\right)\Big]\Big\}.\label{befunction}
\end{align}

The function $I(x,y,z)$ is the $\epsilon^0$ term in the $\epsilon$ expansion of the propagator seagull wih $x_2=x_3=0$, also taken from \cite{Scharf:1993ds} Eq.\ (95).
\begin{align}
&I(x_1,x_4,p^2)= \ln^2(x_4/Q^2)+\Big(-5+\left(1+\frac{x_1/x_4-1}{p^2/x_4}\right)\ln(x_1/x_4)\nonumber \\
&-\frac{r_{41p}-t_{41p}}{p^2/x_4}(\ln(r_{41p})-\ln(t_{41p}))\Big)\ln(x_4/Q^2)+\frac{19}{2}+\frac{3}{2}\zeta(2)+\frac{1}{2}\left(1+\frac{x_1/x_4-1}{p^2/x_4}\right)\ln^2(x_1/x_4)\nonumber \\
&+\left(-2\left(1+\frac{x_1/x_4-1}{p^2/x_4}\right)-\frac{3}{4}\frac{r_{41p}-t_{41p}}{p^2/x_4}(\ln(r_{41p})-\ln(t_{41p}))\right)\ln(x_1/x_4)\nonumber \\
&+\frac{1-p^2/x_4}{p^2/x_4}\ln(1-p^2/x_4)+\frac{1}{2}\left(\frac{1-x_1/x_4+p^2/x_4}{p^2/x_4}\right)\Li_2(p^2/x_4)\nonumber \\
&+\frac{1}{2} \frac{r_{41p}-t_{41p}}{p^2/x_4}\Big[ 4\ln(r_{41p})-4\ln(t_{41p})+\ln\left(\frac{1-r_{41p}}{t_{41p}-r_{41p}}\right)\ln\left(\frac{r_{41p}(1-t_{41p})}{r_{41p}-t_{41p}}\right)\nonumber \\
&-\ln\left(\frac{1-t_{41p}}{r_{41p}-t_{41p}}\right)\ln\left(\frac{t_{41p}(1-r_{41p})}{t_{41p}-r_{41p}}\right)+\Li_2\left(\frac{r_{41p}(1-t_{41p})}{r_{41p}-t_{41p}}\right)-\Li_2\left(\frac{t_{41p}(1-r_{41p})}{t_{41p}-r_{41p}}\right)\nonumber\\
&-\Li_2\left(\frac{1-t_{41p}}{r_{41p}-t_{41p}}\right)+\Li_2\left(\frac{1-r_{41p}}{t_{41p}-r_{41p}}\right)-\Li_2(1-r_{41p})+\Li_2(1-t_{41p})\nonumber\\
&-\Li_2\left(\frac{t_{41p}(1-r_{41p})}{-r_{41p}}\right)-\eta(1-p^2/x_4,1/r_{41p})\ln\left(\frac{t_{41p}(1-r_{41p})}{-r_{41p}}\right)\nonumber\\
&+\Li_2\left(\frac{r_{41p}(1-t_{41p})}{-t_{41p}}\right)+\eta(1-p^2/x_4,1/t_{41p})\ln\left(\frac{r_{41p}(1-t_{41p})}{-t_{41p}}\right)\label{ifunction}
\end{align}
where $\eta(a,b)=\ln(ab)-\ln(a)-\ln(b)$.
\section{Relation of SFI with IBP and DE}
\label{sec:relation}

Here we provide more details on the relation of Symmetries of Feynman Integrals (SFI) with the standard methods of Integration By Parts (IBP) and Differential Equations (DE).

Both IBP and DE are based on roughly the same variations and so is SFI, namely the infinitesimal variation of the loop currents (\ref{trans1},\ref{trans3}) or ``freedom of loop currents''. In fact, SFI interprets the mass variables $x$ also as formal variables of a generating function, which encodes the integrals for all indices $\nu$ (powers of propagators) through derivatives (see \cite{SFI} eq. (2.2-3)). From this point of view the $x$'s and $\nu$'s are conjugate variables related by a transform which also exchanges the IBP recurrence relations in $\nu$'s and the differential equations of DE in $x$'s. In this sense the SFI suggests a unified view of IBP and DE. 

Moreover, SFI uses the same group structure among IBP variations which was noticed in \cite{LeeGroup} and is known as a ``Lee group''. 

At the same time, SFI introduces several novelties: \bi 

\item  SFI identifies the condition for a variation to produce a differential equation for the integral $I$ (with no appearance of extra irreducible scalar products), thereby leading to the definition of the SFI equation system and the associated SFI group $G$.

\item SFI recognizes the action of $G$ on the parameter space $X$ and its foliation into $G$-orbits. After some normalization of the integral $I$ (through division by the homogeneous solution $I_0$, related to the maximal cut) it reduces within each orbit to its value at a conveniently chosen base point plus a line integral over diagrams with one propagator contracted (and hence simpler). The foliation of $X$ and the mentioned reduction are the main point of SFI. As far as we know the theory for a generic reduction to a line integral over simpler diagrams is novel. The word symmetry appearing in the method's name refers to the group $G$, which is a symmetry of $I_0$ and determines the foliation. 

\item SFI identifies the singular locus of the equation system. On this locus the integral can be expressed as a linear combination of simpler diagrams, rather than a line integral thereof. Methods were developed to determine the coefficients of these linear combinations.
\ei

We take this opportunity to discuss some shortcomings of the standard term ``integration by parts'' \bi

\item Indeed the IBP recurrence relations are derived by applying the elementary method of integration by parts. However, the term hides the intimate connection between the equations and the diagram topology and may create the misleading impression that it is nothing more than integration by parts of first year calculus.

\item The SFI equation system includes equations from variations of the form $p \del_p$ where $p$ is an external momentum, which are not total derivatives, namely, they are not associated with integration by parts. 

\item There are IBP  generators which do not produce differential equations for $I$ since they generate irreducible scalar products. 
\ei

\bibliographystyle{unsrt} 

\begin{thebibliography}{99}

\bibitem{CoreTheory}
F.~Wilczek, 
 ``The Lightness of Being'', Basic Books (2010); \\
    ``Physics in 100 Years,''
  Phys.\ Today {\bf 69}, 0432 (2016)
  doi:10.1063/PT.3.3137
  [arXiv:1503.07735 [physics.pop-ph]].

\bibitem{SFI}
  B.~Kol,
  ``Symmetries of Feynman integrals and the Integration By Parts method,'' arXiv:1507.01359 [hep-th].

\bibitem{locus}
  B.~Kol,
  ``The algebraic locus of Feynman integrals,'' arXiv:1604.07827 [hep-th].

\bibitem{bubble}
  B.~Kol, ``Bubble diagram through the Symmetries of Feynman Integrals method,'' arXiv:1606.09257 [hep-th].
  
\bibitem{VacuumSeagull}
  P.~Burda, B.~Kol and R.~Shir, ``Vacuum seagull: Evaluating a three-loop Feynman diagram with three mass scales,'' Phys.\ Rev.\ D {\bf 96} (2017) no.12,  125013 doi:10.1103/PhysRevD.96.125013 [arXiv:1704.02187 [hep-th]].
   
\bibitem{minors}
  B.~Kol,
  ``Algebraic aspects of when and how a Feynman diagram reduces to simpler ones,'' arXiv:1804.01175 [hep-th].

\bibitem{diameter} 
  B.~Kol,
  ``Two-loop vacuum diagram through the Symmetries of Feynman Integrals method,''
  arXiv:1807.07471 [hep-th].

\bibitem{kite} 
  B.~Kol and S.~Mazumdar,
  ``Kite diagram through Symmetries of Feynman Integrals,''
  Phys.\ Rev.\ D {\bf 99} (2019) no.4,  045018
  doi:10.1103/PhysRevD.99.045018
  [arXiv:1808.02494 [hep-th]].

  
\bibitem{Chetyrkin:1981qh}
  K.~G.~Chetyrkin and F.~V.~Tkachov,
  ``Integration by Parts: The Algorithm to Calculate beta Functions in 4 Loops,''
  Nucl.\ Phys.\ B {\bf 192} (1981) 159.
  doi:10.1016/0550-3213(81)90199-1
    
\bibitem{Kotikov:1990kg}
  A.~V.~Kotikov,
  ``Differential equations method: New technique for massive Feynman diagrams calculation,''
  Phys.\ Lett.\ B {\bf 254} (1991) 158.
  doi:10.1016/0370-2693(91)90413-K
  
\bibitem{Remiddi:1997ny}
  E.~Remiddi,
  ``Differential equations for Feynman graph amplitudes,''
  Nuovo Cim.\ A {\bf 110} (1997) 1435
  [hep-th/9711188].
  
 \bibitem{GehrmannRemiddi1999} 
  T.~Gehrmann and E.~Remiddi,
  ``Differential equations for two loop four point functions,''
  Nucl.\ Phys.\ B {\bf 580}, 485 (2000)
  [hep-ph/9912329].

  
\bibitem{Caffo:1998yd}
  M.~Caffo, H.~Czyz, S.~Laporta and E.~Remiddi,
  ``Master equations for master amplitudes,''
  Acta Phys.\ Polon.\ B {\bf 29} (1998) 2627
  [hep-th/9807119].

\bibitem{SmirnovBooks} 
  V.~A.~Smirnov,
  ``Feynman integral calculus,''
  Berlin, Germany: Springer (2006). 

  V.~A.~Smirnov,
  ``Analytic tools for Feynman integrals,''
  Springer Tracts Mod.\ Phys.\  {\bf 250}, 1 (2012).


\bibitem{DiamondRule} 
  B.~Ruijl, T.~Ueda and J.~Vermaseren,
  ``The diamond rule for multi-loop Feynman diagrams,''
  Phys.\ Lett.\ B {\bf 746}, 347 (2015)
  doi:10.1016/j.physletb.2015.05.015
  [arXiv:1504.08258 [hep-ph]].
  
    \bibitem{Tarasov1997} 
  O.~V.~Tarasov,
  ``Generalized recurrence relations for two loop propagator integrals with arbitrary masses,''
  Nucl.\ Phys.\ B {\bf 502}, 455 (1997)
  doi:10.1016/S0550-3213(97)00376-3
  [hep-ph/9703319].
  
  \bibitem{Bauberger1994} 
  S.~Bauberger, M.~B\"ohm, G.~Weiglein, F.~A.~Berends and M.~Buza,
  ``Calculation of two-loop self-energies in the electroweak Standard Model,''
  Nucl.\ Phys.\ Proc.\ Suppl.\  {\bf 37B}, no. 2, 95 (1994)
  doi:10.1016/0920-5632(94)90665-3
  [hep-ph/9406404].
  
\bibitem{Davydychev:2000na}
A.~I.~Davydychev and M.~Y.~Kalmykov, ``New results for the epsilon expansion of certain one, two and three loop Feynman diagrams,'' Nucl.\ Phys.\ B {\bf 605} (2001) 266 doi:10.1016/S0550-3213(01)00095-5 [hep-th/0012189]. 

\bibitem{Boos:1990rg}
  E.~E.~Boos and A.~I.~Davydychev,
  ``A Method of evaluating massive Feynman integrals,''
  Theor.\ Math.\ Phys.\  {\bf 89} (1991) 1052
   [Teor.\ Mat.\ Fiz.\  {\bf 89} (1991) 56].
  doi:10.1007/BF01016805
  
  
\bibitem{Scharf:1993ds}
  R.~Scharf and J.~B.~Tausk, ``Scalar two loop integrals for gauge boson selfenergy diagrams with a massless fermion loop,'' Nucl.\ Phys.\ B {\bf 412} (1994) 523. doi:10.1016/0550-3213(94)90391-3 
  
\bibitem{Bauberger:1994by}
S.~Bauberger, F.~A.~Berends, M.~Bohm and M.~Buza, ``Analytical and numerical methods for massive two loop selfenergy diagrams,'' Nucl.\ Phys.\ B {\bf 434} (1995) 383 doi:10.1016/0550-3213(94)00475-T [hep-ph/9409388].
 
\bibitem{Jegerlehner:2003py}
  F.~Jegerlehner and M.~Y.~Kalmykov,
  ``O(alpha alpha(s)) correction to the pole mass of the t quark within the standard model,''
  Nucl.\ Phys.\ B {\bf 676} (2004) 365
  doi:10.1016/j.nuclphysb.2003.10.012
  [hep-ph/0308216].

\bibitem{Jegerlehner:2003sp}
F.~Jegerlehner and M.~Y.~Kalmykov, ``O(alpha alpha(s)) relation between pole- and MS-bar mass of the t quark,'' Acta Phys.\ Polon.\ B {\bf 34} (2003) 5335 [hep-ph/0310361]. 


\bibitem{Bytev:2011ks}
V.~V.~Bytev, M.~Y.~Kalmykov and B.~A.~Kniehl, ``HYPERDIRE, HYPERgeometric functions DIfferential REduction: MATHEMATICA-based packages for differential reduction of generalized hypergeometric functions $_pF_{p-1}$, $F_1$,$F_2$,$F_3$,$F_4$,'' Comput.\ Phys.\ Commun.\  {\bf 184} (2013) 2332 doi:10.1016/j.cpc.2013.05.009 [arXiv:1105.3565 [math-ph]]. 

\bibitem{Gray:1990yh}
N.~Gray, D.~J.~Broadhurst, W.~Grafe and K.~Schilcher, ``Three Loop Relation of Quark (Modified) Ms and Pole Masses,'' Z.\ Phys.\ C {\bf 48} (1990) 673. doi:10.1007/BF01614703 

\bibitem{Martin:2005qm}
  S.~P.~Martin and D.~G.~Robertson, ``TSIL: A Program for the calculation of two-loop self-energy integrals,'' Comput.\ Phys.\ Commun.\  {\bf 174} (2006) 133 doi:10.1016/j.cpc.2005.08.005 [hep-ph/0501132].


\bibitem{Martin:2003it}
S.~P.~Martin, ``Two loop scalar self energies in a general renormalizable theory at leading order in gauge couplings,'' Phys.\ Rev.\ D {\bf 70} (2004) 016005 doi:10.1103/PhysRevD.70.016005 [hep-ph/0312092]. 

\bibitem{Martin:2003qz}
S.~P.~Martin, ``Evaluation of two loop selfenergy basis integrals using differential equations,'' Phys.\ Rev.\ D {\bf 68} (2003) 075002 doi:10.1103/PhysRevD.68.075002 [hep-ph/0307101]. 
 
 
\bibitem{MaximalCut}
R.~N.~Lee and V.~A.~Smirnov,
  ``The Dimensional Recurrence and Analyticity Method for Multicomponent Master Integrals: Using Unitarity Cuts to Construct Homogeneous Solutions,''
  JHEP {\bf 1212}, 104 (2012)
  doi:10.1007/JHEP12(2012)104
  [arXiv:1209.0339 [hep-ph]]. \\
\bibitem{MaximalCut1}
E.~Remiddi and L.~Tancredi,
  ``Differential equations and dispersion relations for Feynman amplitudes. The two-loop massive sunrise and the kite integral,''
  Nucl.\ Phys.\ B {\bf 907}, 400 (2016)
  doi:10.1016/j.nuclphysb.2016.04.013
  [arXiv:1602.01481 [hep-ph]].\\
  \bibitem{MaximalCut2}
   A.~Primo and L.~Tancredi,
  ``On the maximal cut of Feynman integrals and the solution of their differential equations,''
  Nucl.\ Phys.\ B {\bf 916}, 94 (2017)
  doi:10.1016/j.nuclphysb.2016.12.021
  [arXiv:1610.08397 [hep-ph]].\\

  
\bibitem{Smirnov:2014hma}
  A.~V.~Smirnov,
  ``FIRE5: a C++ implementation of Feynman Integral REduction,''
  Comput.\ Phys.\ Commun.\  {\bf 189} (2015) 182
  doi:10.1016/j.cpc.2014.11.024
  [arXiv:1408.2372 [hep-ph]].
  
\bibitem{LeeGroup} 
  R.~N.~Lee,
  ``Group structure of the integration-by-part identities and its application to the reduction of multiloop integrals,''
  JHEP {\bf 0807}, 031 (2008)
  doi:10.1088/1126-6708/2008/07/031
  [arXiv:0804.3008 [hep-ph]].
  
\end{thebibliography}

\end{document}